\documentclass{achemso}
\setkeys{acs}{journal=acscii, manuscript=article}
\usepackage[numbers,super]{natbib}
\usepackage{amsmath,amsthm}
\usepackage{float}
\usepackage{color,epsfig,multirow}
\usepackage{chemfig}
\usepackage{geometry}
\usepackage{tikz}
\usepackage[utf8]{inputenc}
\usepackage[english]{babel}
\usepackage{amssymb}
\usepackage{booktabs}
\usepackage{hyperref}

\newcommand{\RNum}[1]{\uppercase\expandafter{\romannumeral #1\relax}}

\usepackage{bm}
\usepackage{tikz}

\usepackage{xcolor}

\title{Fortuitous Cancellation of Competing Effects in the Free Energy of Water Autoionization: Long-Range Electrostatics, System Size, and Nuclear Quantum Effects}

\author{Samuel Varner}
\altaffiliation{These authors contributed equally to this work: Samuel Varner, Yinan Xu, and Yezhi Jin}
\affiliation{Department of Chemical and Biomolecular Engineering, Tandon School of Engineering, New York University, Brooklyn, New York 11201, USA}

\author{Yinan Xu}
\email{yx3878@nyu.edu}
\altaffiliation{These authors contributed equally to this work: Samuel Varner, Yinan Xu, and Yezhi Jin}
\affiliation{Department of Chemical and Biomolecular Engineering, Tandon School of Engineering, New York University, Brooklyn, New York 11201, USA}

\author{Yezhi Jin}
\altaffiliation{These authors contributed equally to this work: Samuel Varner, Yinan Xu, and Yezhi Jin}
\affiliation{Department of Chemical and Biomolecular Engineering, Tandon School of Engineering, New York University, Brooklyn, New York 11201, USA}

\author{Gustavo R. P\'erez-Lemus}
\affiliation{Department of Chemical and Biomolecular Engineering, Tandon School of Engineering, New York University, Brooklyn, New York 11201, USA}

\author{Pablo F. Zubieta Rico}
\affiliation{Department of Chemical and Biomolecular Engineering, Tandon School of Engineering, New York University, Brooklyn, New York 11201, USA}

\author{Zhiying Yi}
\affiliation{Pritzker School of Molecular Engineering, The University of Chicago, Chicago, Illinois 60637, USA}

\author{Joan M. Montes de Oca}
\affiliation{Materials Science Division, Argonne National Laboratory, Lemont, Illinois 60439, USA}

\author{Paul F. Nealey}
\affiliation{Pritzker School of Molecular Engineering, The University of Chicago, Chicago, Illinois 60637, USA}
\alsoaffiliation{Materials Science Division, Argonne National Laboratory, Lemont, Illinois 60439, USA}

\author{Seth B. Darling}
\affiliation{Pritzker School of Molecular Engineering, The University of Chicago, Chicago, Illinois 60637, USA}
\alsoaffiliation{Chemical Sciences and Engineering Division, Argonne National Laboratory, Lemont, Illinois 60439, USA}

\author{Zhen-Gang Wang}
\affiliation{Division of Chemistry and Chemical Engineering, California Institute of Technology, Pasadena, California 91125, USA}

\author{Juan J. de Pablo}
\email{jjd8110@nyu.edu}
\affiliation{Department of Chemical and Biomolecular Engineering, Tandon School of Engineering, New York University, Brooklyn, New York 11201, USA}
\alsoaffiliation{Department of Computer Science,
Courant Institute of Mathematical Sciences and Department of Physics, New York University, New York City, New York 10012, USA}

\DeclareUnicodeCharacter{2009}{\,}
\begin{document}
\singlespacing

	\begin{abstract}
        
         Few molecular processes are more fundamental to chemistry than the autoionization of water, whose equilibrium constant, $K_\text{w} \approx 10^{-14}$, underlies acid-base equilibria and the chemistry of life. Yet the molecular origins of the free energy cost of forming and separating ions remain incompletely understood. Recent efforts to explain the autoionization process have been limited by the difficulty of modeling its reactivity in sufficient detail. Here we combine machine learning interatomic potentials (MLIPs) with spectral adaptive biasing force enhanced sampling to connect proton transfer and water wire configurations to the underlying free energy landscape. We identify competing influences of long-range electrostatics, system size, temperature, and nuclear quantum motion on the predicted ionization thermodynamics. The free energy surface extends from neutral water to ions separated beyond the solvent-separated ion pair (SSIP). It reveals a transition state in which a proton is shared between the second and third molecules of a three-membered water wire. The contact ion pair is unstable and the SSIP only weakly metastable, so determining $\text{p}K_\text{w}$ requires sampling longer wires. Long-range electrostatics raise the free energy at ion separations beyond 3~\AA. Models with comparable force errors give substantially different free energy profiles in this region. Increasing system size lowers the calculated $\text{p}K_\text{w}$, while nuclear quantum effects bring it closer to experiment and reproduce the higher $\text{p}K_\text{w}$ of $\text{D}_2\text{O}$ relative to $\text{H}_2\text{O}$. By systematically assessing these contributions, we establish a baseline for interpreting thermodynamic results from simulations using MLIPs in which charged and polar groups react and diffuse in finite simulation cells. Our analysis also explains how cancellation among competing effects can yield ionization constants close to experiment even when simulations omit several physical contributions.
	\end{abstract}

    \clearpage
        
    \section{Introduction}
        Water autoionization, the dissociation of water into hydronium (H$_3$O$^+$) and hydroxide (OH$^-$) ions, gives rise to one of the most fundamental constants in chemistry: the ionization constant of water, $K_\text{w} \approx 10^{-14}$ at 298~K. This constant sets the scale of acid-base equilibria in chemical and biological systems, from reactions in human cells to atmospheric chemistry and industrial processes, and in this sense forms part of the chemical basis for life.\cite{zhou_electrostatic_2018,ball_water_2008,ball_water_2017,lund_charge_2005,lund_charge_2013,silva_polyelectrolyte_2009,levy_water_2006,eigen_untersuchungen_1955,vacha_charge_2012,xiong_strong_2020,chamberlayne_simple_2020,hao_can_2022,martins-costa_electrostatics_2023} Over several decades, computational studies have sought to explain the molecular mechanism underlying this equilibrium, yet the mechanistic details of autoionization remain a subject of debate.\cite{trout_analysis_1999,mei_quantum_1998,sato_theoretical_1998,geissler_autoionization_2001,mann_water_2003,yoshida_ab_2006,buch_water_2007,beattie_surface_2008,ceriotti_nuclear_2013,da_silva_revisiting_2013} At room temperature, autoionization involves proton transfer along chains of hydrogen-bonded water molecules, often called water wires, with subsequent ion separation proceeding through the Grotthuss mechanism.\cite{de_grotthuss_memoire_1805} Much remains unknown about the transition states and ionic intermediates, and how temperature and nuclear quantum motion influence these processes. Previous computational studies have suggested a collective process involving multiple water molecules, with asynchronous hydrogen transfers and the formation of compact four-member water wires,\cite{liu_mechanistic_2023,zhang_propensity_2024,dasgupta_nuclear_2025} but the underlying free energy landscape and several reaction intermediates remain poorly resolved.

        Resolving these questions also requires assessing how computational approximations affect the predicted mechanism and thermodynamics. Early studies based on \textit{ab initio} molecular dynamics were restricted to small systems and short timescales, which limited sampling of the free energy landscape. Machine learning interatomic potentials (MLIPs) made larger systems and longer simulations accessible by approximating reference density functional theory (DFT) energies and forces at a fraction of the computational cost.\cite{deringer_machine_2019,wang_machine_2024,jacobs_practical_2025,kalita_machine_2025} However, until recently, most MLIPs used only local atomic environments and could not explicitly represent long-range electrostatic interactions.\cite{zhang_deep_2022,anstine_machine_2023,ji_machine-learning_2025} As a result, most studies limited their analysis of the autoionization pathway to short ion separation distances, where electrostatic interactions between ions were still thought to be largely captured by short-range MLIPs. Additionally, despite the efficiency of modern MLIPs, an overwhelming majority of studies limited their analysis to small boxes containing no more than 256 water molecules. Small systems can introduce finite-size effects even with short-range models, and explicit long-range interactions may add further size dependence. Lastly, nuclear quantum effects such as tunneling and the zero-point energy were largely ignored, despite their documented importance in the thermodynamics and kinetics of proton transfer involving light atoms, even at room temperature.\cite{ceriotti_nuclear_2013,markland_nuclear_2018} The contributions of long-range electrostatics, system size, and nuclear quantum motion must therefore be assessed together rather than in isolation. Several studies have nevertheless reported pK$_w$ values close to experiment despite omitting some or all of these contributions.

        Here, we map the free energy landscape of water autoionization from neutral water to separated ion pairs using MLIPs that explicitly account for long-range electrostatic interactions,\cite{zhang_deep_2022,wang_deepmd-kit_2018,zeng_deepmd-kit_2023} combined with a recently proposed spectral adaptive biasing force (spectral-ABF) enhanced sampling technique.\cite{rico_sobolev_2022,zubieta_rico_pysages_2024,perez-lemus_importance_2024} We combine one-dimensional sampling of the initial proton transfer with two-dimensional sampling in the two-ion region to resolve the free energy surface (FES) and minimum free energy pathway (MFEP). We also characterize the equilibrium distribution of water wire configurations along this pathway. We systematically assess the roles of long-range electrostatic interactions, system size, temperature, and nuclear quantum effects on the autoionization free energy and pK$_w$. The predicted free energies and ionization constants are sensitive to each of these factors. Within the models examined here, small system sizes and neglected nuclear quantum effects both raise the product free energy relative to neutral water, while missing long-range Coulomb interactions lower it. These opposing effects can partially cancel, giving pK$_w$ values close to experiment for particular combinations of model and box size. Including these effects together brings the calculated pK$_w$ closer to experiment and reproduces the isotopic ordering pK$_w$($\text{D}_2\text{O}$) $>$ pK$_w$($\text{H}_2\text{O}$).\cite{bandura_ionization_2006,lide_crc_2005}
        
        At 300~K, the free energy profile corresponding to the Grotthuss mechanism\cite{de_grotthuss_memoire_1805,agmon_grotthuss_1995} reveals a primary transition state marked by a three-membered water wire in which the proton is equally shared between the second and third water molecules. This transition state leads to a shallow metastable basin corresponding to the solvent-separated ion pair (SSIP), characterized by two fully formed autoions separated by a single water molecule, with a recombination barrier below $k_BT$. A smaller barrier ($\sim 2k_BT$) must then be overcome before the free energy plateaus at ion separations corresponding to four-membered and longer water wires. These observations are consistent with transition path sampling simulations conducted by Moqadam \textit{et al.}\cite{moqadam_local_2018} Additionally, we find that the hydroxide ion is hypercoordinated throughout the autoionization pathway, with slightly more than four hydrogen bonds on average, in line with previous observations by Hassanali \textit{et al.}\cite{hassanali_recombination_2011} These results show why sampling must extend to four-membered and longer water wires, where long-range electrostatics and system size substantially affect the free energy.

    \section{Results and Discussion}

    \subsection{Autoionization free energy landscape and $\mathrm{p}K_\mathrm{w}$}
    \label{sec:results-fe}

    We use two machine learning interatomic potentials trained on the same revPBE-D3 reference data: a short-range Deep Potential model (DPSR),\cite{zhang_end--end_2018} and a model with explicit long-range electrostatics (DPLR) based on a deep Wannier description of molecular polarization.\cite{zhang_deep_2022} We sample two collective variables (CVs): $\text{CV}_\text{num}$, the effective number of autoions, and $\text{CV}_\text{dist}$, their effective separation. $\text{CV}\text{num}$ provides a continuous measure of ion formation by quantifying deviations from two hydrogens per oxygen. $\text{CV}\text{dist}$ uses smooth proton assignments to track the separation of hydronium and hydroxide character as protons move through the water network. Spectral-ABF sampling\cite{zubieta_rico_efficient_2025} in PySAGES\cite{zubieta_rico_pysages_2024} is performed along $\text{CV}_\text{num}$ during the initial proton transfer and along both CVs in the two-ion region. Nuclear quantum effects are included using path integral molecular dynamics with the PIGLET thermostat.\cite{ceriotti_accelerating_2011,ceriotti_efficient_2012} The Supporting Information gives the CP2K parameters,\cite{kuhne_cp2k_2020} CV definitions, model architectures and validation, sampling and reweighting procedures, and path integral setup.

    We begin by presenting the autoionization free energy computed using the long-range DPLR model with 512 water molecules at 300~K. This setup serves as our reference throughout the paper, against which we later assess the impact of model choice, system size, temperature, and nuclear quantum effects.
    
    \begin{figure}[!h]
        \centering
        \includegraphics[width=1.0\linewidth]{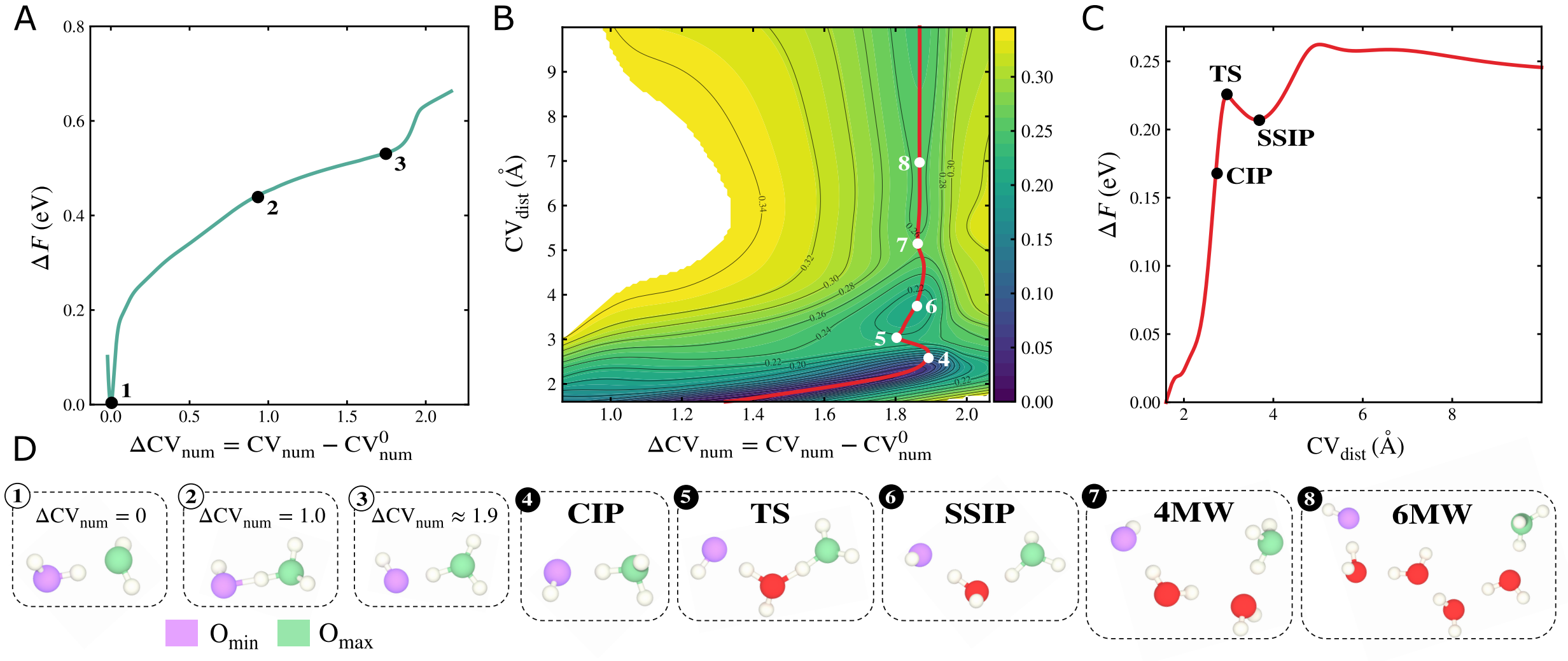}
        \caption{(A) One-dimensional free energy profile for water autoionization computed from spectral-ABF sampling in $\text{CV}_\text{num}$ with the DPLR model and 512 water molecules at 300~K. $\Delta\text{CV}_\text{num}$ denotes that the profile has been shifted to place the neutral-water minimum at zero. (B) Two-dimensional free energy surface for water autoionization (eV units), beginning near the end of the initial proton transfer region. The minimum free energy pathway is shown as a red line, with numbered white markers corresponding to configurations 4-8 in panel (D). (C) The minimum free energy path projected onto $\text{CV}_\text{dist}$ and shifted to zero reference. (D) Representative configurations along the autoionization pathway, labeled in panels (A) and (B). Abbreviations denote the contact ion pair (CIP), transition state (TS), solvent-separated ion pair (SSIP), four-membered wire (4MW), and six-membered wire (6MW).}
        \label{fig:1d-fep}
    \end{figure}
    
    Figure~\ref{fig:1d-fep}A shows the one-dimensional free energy profile obtained from spectral-ABF sampling along $\text{CV}_\text{num}$. The profile features a minimum corresponding to the value of all rational switching functions at the equilibrium O-H bond distance, denoted CV$_\text{num}^0$. We shift the profile horizontally by CV$_\text{num}^0$ so that this minimum sits at $\Delta\text{CV}_\text{num} = 0$, since the unshifted baseline depends weakly on the system size and would otherwise obscure direct comparison between cell sizes. Characteristic configurations along the profile (Figure~\ref{fig:1d-fep}D) show that the system passes from the neutral state through a water-dimer (WD) configuration, in which two adjacent molecules evenly share a hydrogen, toward a contact ion pair (CIP), in which a fully formed $\text{H}_3\text{O}^+$ and $\text{OH}^-$ ions are in direct hydrogen-bonded contact. Sampling along $\text{CV}_\text{num}$ alone does not resolve ion separation near the two-ion state. The profiles and configurations in Figure~\ref{fig:1d-fep}A, B, and D show that points 3--8 have similar values of $\text{CV}_\text{num}$ despite their different ion separations and water wire configurations.

    To resolve this orthogonal direction, we perform two-dimensional spectral-ABF along both CV$_\text{num}$ and CV$_\text{dist}$ in the two-ion region, as described in the Supporting Information. Figure~\ref{fig:1d-fep}B shows the resulting free energy surface (FES) together with the minimum free energy path (MFEP) obtained using the string method. The MFEP projected onto $\text{CV}_\text{dist}$ is shown in Figure~\ref{fig:1d-fep}C. The CIP state is shown as configuration 4 in Figure~\ref{fig:1d-fep}D. This is followed by a barrier crossing in the transition state (TS), beyond which lies a shallow basin associated with the solvent-separated ion pair (SSIP) and a long plateau corresponding to extended water-wire configurations. The TS is characterized by a three-membered water wire with the proton shared evenly between the second and third water molecules, as shown in Figure~\ref{fig:1d-fep}D. Interestingly, even though the SSIP contains two fully formed ions separated by an intervening water molecule, the barrier to recombination in the SSIP is less than the thermal energy $k_B T$. The SSIP therefore offers little free energy resistance to recombination. Free energy calculations that terminate at the CIP or SSIP therefore miss a key part of the reaction mechanism, namely the separation of the ions to a distance where the free energy truly plateaus.
    
    The three-membered transition state and the shallow metastability of the SSIP state agree with transition path sampling simulations by Moqadam \textit{et al.}, who found that states with ions separated by two water molecules lived significantly longer than those separated by one. Thus, three proton transfer events are required before the ionized state can persist.\cite{moqadam_local_2018} Similarly, they found that the first and second transfer events must occur in a concerted manner, aligning with our observation that the transition state occurs during the second proton transfer. They also noted that the third proton transfer can occur sequentially, which is consistent with the small additional barrier ($\sim 2k_B T$) that the system must overcome to proceed from the SSIP state to four-membered and longer water wires. Beyond the SSIP, the ions separate via the Grotthuss mechanism (proton diffusion along the water wire),\cite{de_grotthuss_memoire_1805,agmon_grotthuss_1995} and the free energy effectively plateaus past the four-membered wire (4MW) configuration. While not explicitly studied here, further extension of the wire is expected to proceed via the standard proton diffusion mechanism. The barriers for these subsequent steps are expected to be comparable to those reported by Gomez \textit{et al.},\cite{gomez_neural-network-based_2024} which are small relative to the overall autoionization free energy.
    

    To obtain a full free energy profile that spans the entire reaction, we combine the one-dimensional profile in the initial transfer region, reweighted from $\text{CV}_\text{num}$ onto $\text{CV}_\text{dist}$, with the two-dimensional MFEP, following the procedure described in the Supporting Information and depicted in Figure~S9. The two profiles join continuously at $\text{CV}_\text{dist} \approx 2$~\AA. We then evaluate the reaction free energy from\cite{dietschreit_how_2022}
    \begin{equation}\label{eq:rxn-fe}
        \Delta \mathcal{F} = -k_B T \ln\!\left\{\frac{\int_P d\xi\, e^{-\beta \Delta F(\xi)}}{\int_R d\xi\, e^{-\beta \Delta F(\xi)}}\right\},
    \end{equation}
    where $\xi = \text{CV}_\text{dist}$, $\beta = 1/(k_B T)$, and $R$ and $P$ denote the reactant and product regions. The reactant region is taken as $0 < \text{CV}_\text{dist} < 0.5$~{\AA}, and the product region is taken as $6 < \text{CV}_\text{dist} < 6.5$~{\AA}, which corresponds to states where the ions are separated by two or three water molecules. We note that the absolute value of pK$_w$ is sensitive to the explicit choice of the integration region, since the product region is a plateau and not a well-defined basin. As a result, we employ the same integration bounds for all pK$_w$ calculations to ensure that trends result from the underlying methodological choices and not the arbitrary choice of integration limits. The main purpose of this work is to evaluate the effects and trends resulting from various methodological choices, and the partition ratio method is a common way to evaluate reaction free energies that has been used previoulsy in studies of water autoionization. \cite{dasgupta_nuclear_2025} The corresponding $\text{p}K_w$ is given by
    \begin{equation}\label{eq:pkw}
        \frac{1}{2}\text{p}K_w = \frac{\Delta \mathcal{F}}{2.303\, k_B T} - \log_{10}\!\left(\frac{N_w}{V c_0}\right),
    \end{equation}
    where $N_w$ is the number of water molecules, $V$ is the cell volume, and $c_0 = 6.0221 \times 10^{-4}~\text{\AA}^{-3}$ is the standard concentration of 1~M. Briefly, the first term is the reaction free energy from Equation~\eqref{eq:rxn-fe}, while the second term is an entropic correction that accounts for the standard state reference.\cite{liu_mechanistic_2023,calegari_andrade_probing_2023,dasgupta_nuclear_2025}
    
    Applying Equation~\eqref{eq:pkw} to the resulting profile yields $\text{p}K_w = 18.40$, which is substantially higher than the experimental values of 14.00 for $\text{H}_2\text{O}$ and 14.87 for $\text{D}_2\text{O}$ at 298~K.\cite{lide_crc_2005,bandura_ionization_2006} We next examine the physical origins of this discrepancy. In the following subsection, we systematically vary the treatment of long-range electrostatics, the system size, and the temperature, each of which has a substantial effect on the calculated $\text{p}K_w$. We then assess the role of nuclear quantum effects, which previous work has shown to substantially decrease the computed $\text{p}K_w$.\cite{dasgupta_nuclear_2025} The approach described above, and in detail in the Supporting Information, is used throughout the remainder of the work.

    \subsection{Consequences of long-range interactions, system size, and temperature on $\mathrm{p}K_\mathrm{w}$}

    Two recent studies illustrate the practical importance of these methodological choices. Dasgupta et al.,\cite{dasgupta_nuclear_2025} using a short-range MLIP, reported a free energy profile that is largely insensitive to system size. Calegari Andrade et al.,\cite{calegari_andrade_probing_2023} who explicitly included long-range electrostatics, found instead that the profile depends strongly on the size of the system. We compare both models across system sizes to examine these differences. We also test whether the simulations reproduce the experimental temperature dependence. We begin with the treatment of long-range electrostatic interactions.
    
    Most MLIPs write the total energy as a sum of atomic contributions that depend only on neighbors within a finite cutoff. This local description can reproduce much of the structure of neutral water, but it does not explicitly represent the long-range Coulomb interaction. That limitation becomes particularly relevant when hydronium and hydroxide form and separate, changing both the charge distribution and the polarization of the surrounding water. DPLR adds an electrostatic term based on charges at the nuclei and predicted Wannier centroids, evaluated by Ewald summation under periodic boundary conditions.\cite{zhang_deep_2022} We compare it with DPSR to assess how this treatment affects the autoionization free energy.
    
    \begin{figure}[h!]
        \centering
        \includegraphics[width=1.0\linewidth]{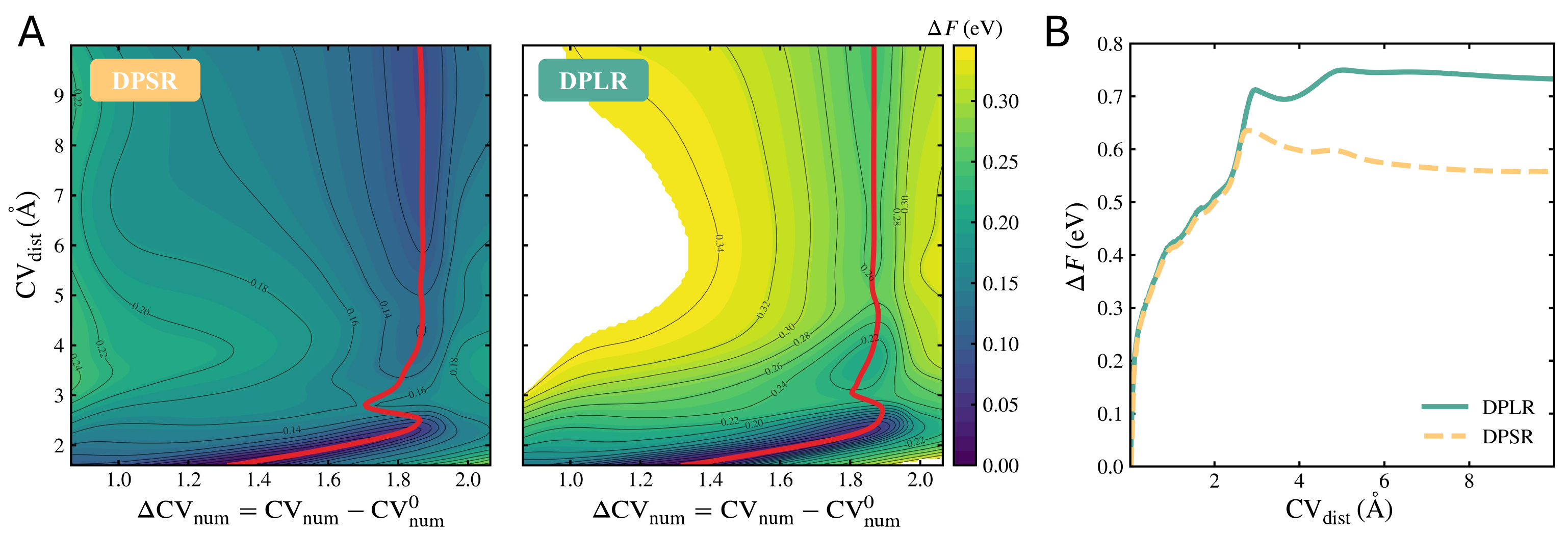}
        \caption{Model dependence of the autoionization free energy for 512 water molecules at 300~K. (A) Two-dimensional free energy surfaces computed with short-range (DPSR) and long-range (DPLR) MLIPs. The color bar applies to both surfaces. Red lines indicate the minimum free energy pathways computed using the string method. (B) Full one-dimensional free energy profiles along $\text{CV}_\text{dist}$, constructed by combining the one-dimensional and two-dimensional sampling regions as described in the Supporting Information.}
        \label{fig:sr-vs-lr}
    \end{figure}
    
    In Figure~\ref{fig:sr-vs-lr}, we compare the free energy surfaces and combined one-dimensional profiles from the DPSR and DPLR models. The DPSR surface in Figure~\ref{fig:sr-vs-lr}A has a lower overall free energy and shows a substantial decrease after the transition state, relative to the DPLR surface. This trend carries over to the overall profiles in Figure~\ref{fig:sr-vs-lr}B, where the two curves are nearly identical at short ion distances ($\text{CV}_\text{dist} < 2$ \AA) and diverge at larger distances. The onset of this divergence near $\text{CV}_\text{dist} \approx 3$ \AA{} is notable because it occurs well within the DPSR descriptor cutoff of 6 \AA. The divergence therefore begins before the ions move beyond the descriptor cutoff. Their separation also changes the surrounding solvent polarization, which extends beyond either local environment. The two models have similar force errors against the same DFT reference, as shown in the Supporting Information, yet give substantially different free energy profiles once the ions form. Since many previous studies employ short-range MLIPs in small simulation cells, we next examine the combined effects of model choice (short versus long) and system size.
    
    \begin{figure}[!h]
        \centering
        \includegraphics[width=0.85\linewidth]{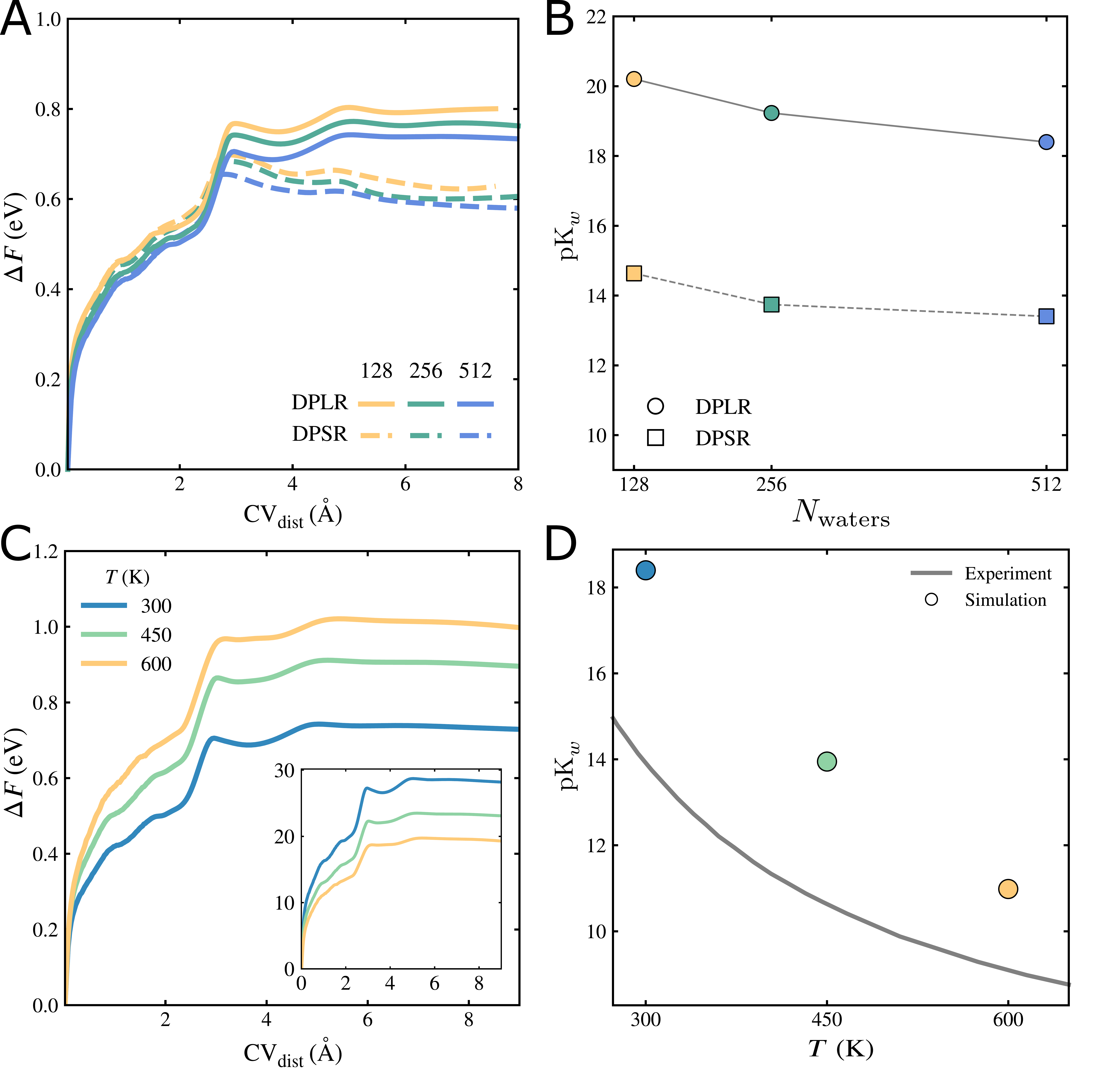}
        \caption{System size and temperature dependence of water autoionization. (A, B) Short-range (DPSR) and long-range (DPLR) models evaluated at various system sizes, all at 300~K. (C, D) DPLR with 512 water molecules evaluated at various temperatures. All simulations use a fixed density of 1.0~g~cm$^{-3}$. (A) Full one-dimensional free energy profiles along $\text{CV}_\text{dist}$, constructed by combining the one-dimensional and two-dimensional sampling regions as described in the Supporting Information. (B) $\text{p}K_w$ values computed from panel (A). (C) Full one-dimensional free energy profiles along $\text{CV}_\text{dist}$ computed at different temperatures. The inset shows the free energy profiles normalized by the thermal energy, $k_BT$. (D) Temperature dependence of $\text{p}K_w$. Reference $\text{p}K_w$ values at a fixed density of 1 g/mL are taken from the correlation shown in Figure 6 of Bandura and Lvov.\cite{bandura_ionization_2006}}
        \label{fig:size-effect}
    \end{figure}
    
    The free energy profiles from one-dimensional sampling purely along $\text{CV}_\text{num}$ decrease with system size for both models, while the results for the two models remain close until the two-ion state (Figure~S10 in the Supporting Information). DPSR then develops a small minimum at the two-ion state that is absent in DPLR. This is consistent with its lower free energy at larger ion separations. Biasing the ion count alone can leave ion separation poorly sampled and obscure this difference.
    
    Figure~\ref{fig:size-effect}A shows the full free energy profiles along $\text{CV}_\text{dist}$, and Figure~\ref{fig:size-effect}B gives the corresponding $\text{p}K_w$. Both models show a decrease in $\text{p}K_w$ with increasing system size, with a larger change for DPLR. DPLR also gives higher $\text{p}K_w$ values at every size studied. The shared size dependence may reflect constraints on solvent orientation and the hydrogen-bond network in small periodic cells. DPLR additionally represents electrostatic interactions with periodic images explicitly. Both models are trained on periodic DFT data, but DPLR separates the long-range contribution from the local energy during training. DPSR must fit the total reference energy using only local environments. These differences offer possible explanations for the size trends, although the present comparison does not separate their individual contributions.

    The models also differ in their dielectric response. DPLR gives $\varepsilon_r \approx 90$ with little size dependence, whereas DPSR gives larger values (Figure~S8 in the Supporting Information). Stronger screening could contribute to the lower separation free energy in DPSR, but the dielectric measurements alone do not quantify that contribution.
    
    The most striking feature of Figure~\ref{fig:size-effect}B is that DPSR with 128 water molecules yields a $\text{p}K_w$ of 14.64, which appears to be in close agreement with the experimental value of 14.00 for $\text{H}_2\text{O}$. This agreement is coincidental. The $\text{p}K_w$ is elevated by the small system size and the neglect of nuclear quantum effects, and depressed by the absence of long-range Coulomb interactions, and these errors happen to cancel for this particular combination of model, box size, and temperature.

    Having established the role of long-range interactions and system size, we now examine the temperature dependence of the autoionization free energy and the resulting $\text{p}K_w$ using the DPLR model with 512 water molecules. Figure~\ref{fig:size-effect}C shows the full one-dimensional free energy profiles along $\text{CV}_\text{dist}$ computed at temperatures ranging from 300 to 600 K. The reaction free energy increases with increasing temperature, but when normalized by the thermal energy $k_B T$ (inset of Figure~\ref{fig:size-effect}C) the profiles decrease with temperature. Since equilibrium populations are governed by $\Delta F / k_B T$, this implies that the autoionization equilibrium shifts toward the ionized state at higher temperatures. This is further reflected in the values of $\text{p}K_w$ in Figure~\ref{fig:size-effect}D, which decrease monotonically with temperature, qualitatively in agreement with the experimental trend for liquid water.\cite{bandura_ionization_2006} The classical DPLR result more closely approaches the experimental $\text{p}K_w$ as temperature increases, consistent with the expectation that nuclear quantum effects become less important at higher temperatures.\cite{ceriotti_efficient_2012,ceriotti_nuclear_2013,ceriotti_nuclear_2016,markland_nuclear_2018} While the recombination barrier from the SSIP state remains small across all temperatures studied, the SSIP basin appears to deepen at lower temperatures. Characterizing how this metastability evolves as the temperature decreases below room temperature and approaches freezing would be an intriguing direction for future work.

    \subsection{Hydrogen-bonded water wire structure}

    \begin{figure}[!h]
        \centering
        \includegraphics[width=1.0\linewidth]{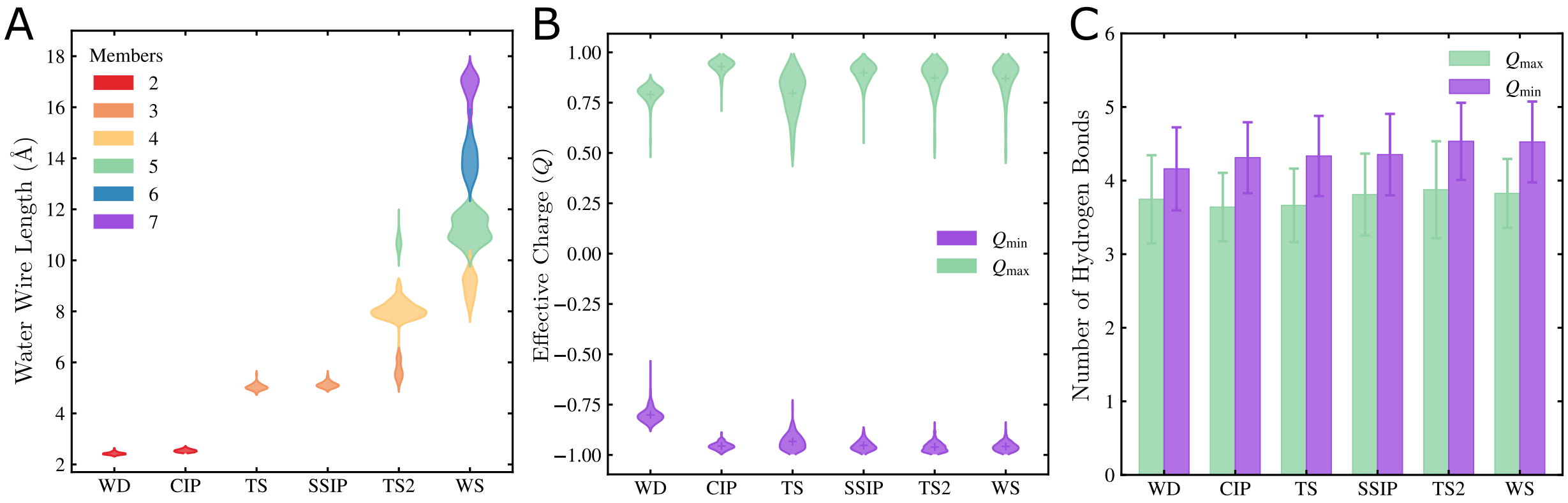}
        \caption{Water wire statistics for the DPLR model with 256 water molecules at 300~K. States are the water dimer (WD), contact ion pair (CIP), transition state (TS), solvent-separated ion pair (SSIP), secondary barrier (TS2), and well-separated ion pair (WS). (A) Distributions of water wire lengths, colored by the number of participating molecules. (B) Distributions of the minimum and maximum effective oxygen charges. (C) Mean hydrogen-bond coordination numbers of the oxygens with minimum and maximum effective charge.}
        \label{fig:water-wire}
    \end{figure}

    It is well known that proton transfer in water, including during water autoionization, proceeds primarily through the Grotthuss mechanism.\cite{de_grotthuss_memoire_1805,agmon_grotthuss_1995} To study the propagation of the proton through the solvent, we define the water wire as the shortest hydrogen-bonded path connecting the hydroxide and hydronium oxygens. Hydroxyl and hydronium oxygens are identified as those with the lowest effective charges ($Q_\text{min}$) and highest effective charges ($Q_\text{max}$), respectively, according to the effective charge $q_i$ of Equation~S4 in the Supporting Information. These effective charges describe proton coordination relative to neutral water, rather than physical electrostatic charges. In Figure \ref{fig:water-wire}A, we plot the length of the water wire at different key states along the autoionization pathway. The violin plots for each state are colored according to the number of water molecules participating in the wire. The water wire length is at least the straight-line distance between the two reactive oxygens and is generally longer for wires with more than two members. Unsurprisingly, the number of water molecules and the length of the water wire are highly correlated. Wires containing more molecules exhibit a larger spread of possible distances as a result of the greater number of available paths and the plateau of the O-O radial distribution function (see Figure~S7 in the Supporting Information). In contrast, states along the MFEP with relatively larger separation distances ($\text{CV}_\text{dist}$) can support water wires with a broader distribution of participating molecules. The water dimer (WD) and contact ion pair (CIP) states are both characterized solely by two-membered water wires, as expected. In the first transition state, the vast majority of water wire configurations have three members. As shown in Figures \ref{fig:1d-fep}C and \ref{fig:1d-fep}D, the three-membered transition state features the transfer of the proton between the second and third water molecules, connecting the CIP and SSIP states. We also provide the distribution of the effective charge states in the hydroxide and hydronium molecules at key points along the MFEP in Figure \ref{fig:water-wire}B. In the three-member transition state, both $Q_\text{min}$ and $Q_\text{max}$ show an increased spread and a skew toward lower magnitudes. The decrease in $Q_\text{max}$ occurs primarily due to the sharing of the proton between the second and third water molecules. Interestingly, Figure~\ref{fig:water-wire}B shows that hydronium takes on a wider range of effective charges at each state along the reaction pathway, indicating that the proton excess is more readily shared with the surrounding solvent. This is consistent with the extension of the water wire from the hydronium end. Additionally, we find that the minimally charged molecule (pre-hydroxide and hydroxide) remains hypercoordinated throughout the autoionization process, averaging slightly more than four hydrogen bonds, as shown in Figure \ref{fig:water-wire}C. This is consistent with the observations of Hassanali \textit{et al.}, who found that hydroxide hypercoordination could play a role in promoting autoionization.\cite{hassanali_recombination_2011,moqadam_local_2018}

    \subsection{Nuclear quantum effects significantly lower $\mathrm{p}K_\mathrm{w}$}

    \begin{figure}[h!]
        \centering
        \includegraphics[width=1.0\linewidth]{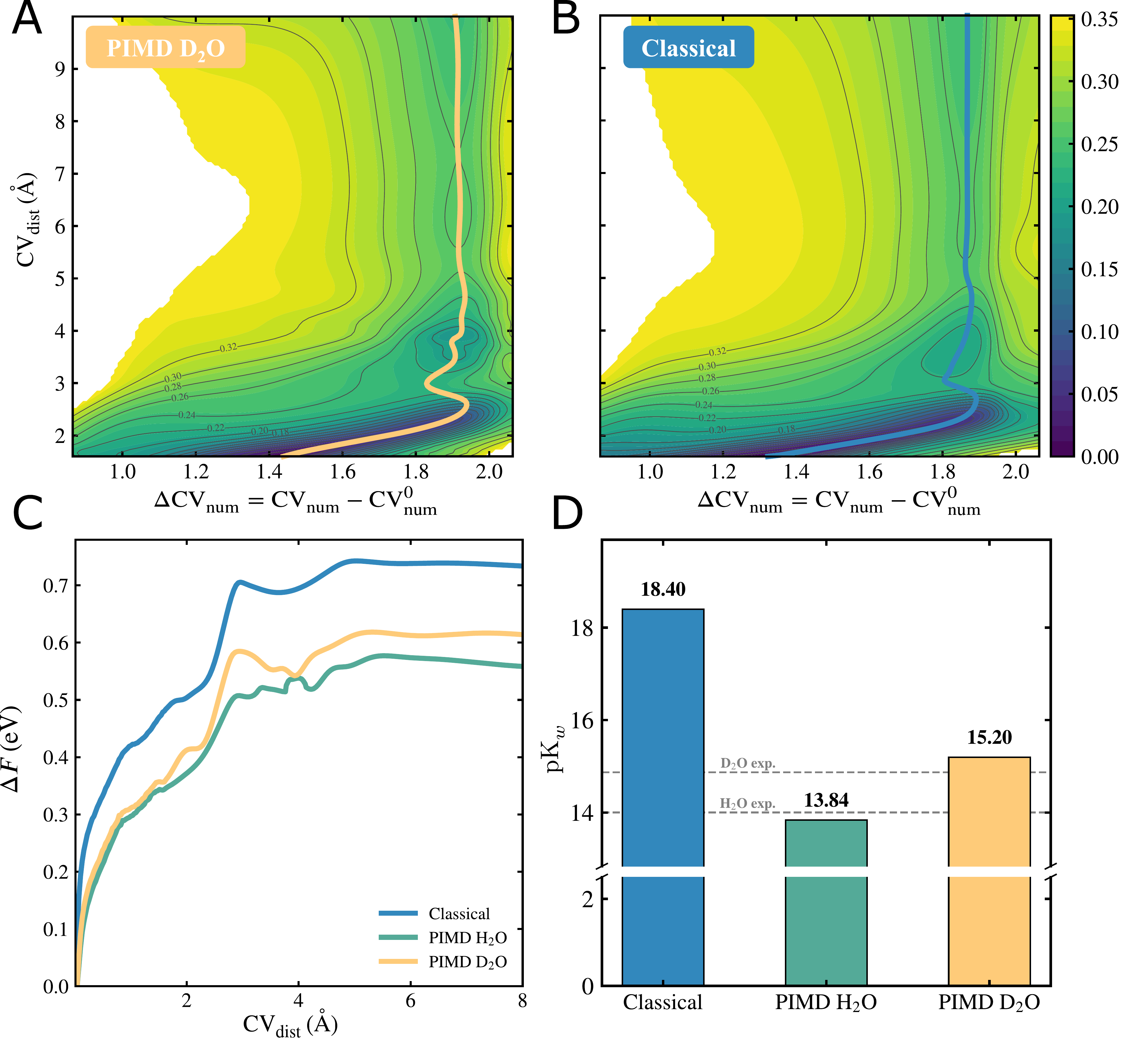}
        \caption{Autoionization free energy comparison between classical MD and path integral molecular dynamics (PIMD) for 512 $\text{H}_2\text{O}$ or $\text{D}_2\text{O}$ molecules using DPLR at 300~K. (A) two-dimensional FES for $\text{D}_2\text{O}$ from PIMD simulations. (B) two-dimensional FES from classical MD simulations. The shared color scale in (A, B) is in eV, and the overlaid lines are the MFEPs. (C) Full autoionization free energy profiles for classical $\text{H}_2\text{O}$, quantum $\text{H}_2\text{O}$, and quantum $\text{D}_2\text{O}$. (D) $\text{p}K_w$ values computed from the free energy profiles in (C). The experimental $\text{p}K_w$ of $\text{H}_2\text{O}$ at 298~K is 14.00,\cite{bandura_ionization_2006} and that of $\text{D}_2\text{O}$ at 298~K is 14.87.\cite{lide_crc_2005}}
        \label{fig:quantum}
    \end{figure}

    Among the most commonly neglected effects in free energy calculations of chemical reactions from molecular simulation are nuclear quantum effects (NQEs). In classical simulations, the nuclei are treated as point particles with no quantum character, so contributions from zero-point energy (ZPE) and quantum tunneling are absent. For reactions involving light atoms, these contributions are far from negligible, even at room temperature.\cite{ceriotti_nuclear_2013,ceriotti_nuclear_2016,markland_nuclear_2018} In water, the high frequency of O-H stretching modes makes their ZPE substantial. Quantum nuclear motion changes the relative free energies of neutral and ionized states and can therefore shift the equilibrium constant.

    To include NQEs, we use path integral molecular dynamics (PIMD), in which each nucleus is represented by a ring of $P$ classical replicas (beads) that recovers the quantum partition function in the limit $P \to \infty$.\cite{kapil_i-pi_2019} We combine DPLR, spectral-ABF, and the PIGLET thermostat to sample the quantum free energy surface. Full details are provided in the Supporting Information. The PIMD simulations are conducted at 300~K for boxes of 512 $\text{H}_2\text{O}$ or $\text{D}_2\text{O}$ molecules. One of the most direct manifestations of NQEs is the isotope effect, where different isotopes yield different thermodynamic and kinetic properties, in contrast to classical simulations, where the particle mass has no effect on equilibrium configurational properties at fixed number density. NQEs are generally weaker for heavier atoms and at higher temperatures, so we expect the difference between classical and quantum results to be more pronounced for $\text{H}_2\text{O}$ than for $\text{D}_2\text{O}$. We apply the same enhanced sampling framework as in the classical simulations to compute the one-dimensional free energy profiles, two-dimensional free energy surfaces, and the resulting $\text{p}K_w$ from PIMD, and compare with the classical results in Figure~\ref{fig:quantum}.
    
    The two-dimensional free energy surfaces in Figures~\ref{fig:quantum}A and \ref{fig:quantum}B show that the qualitative autoionization pathway is relatively insensitive to NQEs. One notable difference is that the two-ion state is shifted to slightly higher $\text{CV}_\text{num}$ in the quantum simulations compared to the classical case. This shift is consistent with the broader distribution of the lengths of the O-H bonds induced by ZPE,\cite{hassanali_recombination_2011,ceriotti_nuclear_2013,markland_nuclear_2018} which alters the average value of the switching functions that define $\text{CV}_\text{num}$. Figure~\ref{fig:quantum}C shows that including NQEs leads to a substantial decrease in the free energy barrier, with quantum $\text{H}_2\text{O}$ exhibiting a barrier lower than quantum $\text{D}_2\text{O}$, consistent with the weaker nuclear quantum effects of the heavier isotope. A similar trend was reported in the recent work of Dasgupta et al. \cite{dasgupta_nuclear_2025}. As shown in Figure~\ref{fig:quantum}D, the calculated $\text{p}K_w$ approaches the experimental value for both isotopes only when NQEs are explicitly included, and the simulations also reproduce the experimental ordering $\text{p}K_w(\text{D}_2\text{O}) > \text{p}K_w(\text{H}_2\text{O})$. The quantum values are 13.84 for $\text{H}_2\text{O}$ and 15.20 for $\text{D}_2\text{O}$, compared with 18.40 classically. The predicted isotope difference is therefore 1.36 pK$_w$ units, larger than the experimental difference of 0.87 at 298~K. Nuclear quantum effects recover the isotope ordering and substantially improve agreement with experiment, although the isotope difference remains overestimated.
    
    \section{Conclusion}
    By combining MLIPs with spectral-ABF sampling, we mapped the free energy landscape of water autoionization at the revPBE-D3 level. The two-dimensional surface resolves a transition state with a three-membered water wire, a weakly metastable SSIP with a recombination barrier below $k_B T$, and a secondary barrier of approximately $2k_B T$ leading to longer wires. Hydroxide remains hypercoordinated along the pathway, with slightly more than four hydrogen bonds on average.\cite{hassanali_recombination_2011,moqadam_local_2018} The landscape is consistent with earlier path sampling results in which the first two proton transfers are concerted and a third permits longer-lived ion separation.\cite{moqadam_local_2018} Sampling must therefore extend beyond the CIP and SSIP to characterize the separated ions.

    Long-range electrostatics, system size, and nuclear quantum effects all change the calculated pK$_w$. DPLR raises the product free energy relative to DPSR, while larger cells and quantum nuclear motion lower it. Their opposing effects explain how short-range models in small cells can give values close to experiment despite omitted physics. Including NQEs with DPLR brings the calculated pK$_w$ closer to experiment for both isotopes and recovers their ordering, although the isotope difference is larger than observed. The remaining size dependence and sensitivity to the product integration bounds also warrant care when interpreting absolute pK$_w$ values.

    These findings are relevant to other aqueous reactions involving proton transfer and charge separation, including acid dissociation, ion-pair separation in electrolytes, and proton transport through confined water networks. In such systems, resolving local bond rearrangements may be insufficient to describe the free energy of the separated products. Our comparison also shows that similar force errors against a common DFT reference do not ensure similar reaction free energies. Testing the free energies and solvent response relevant to the reaction is therefore an essential part of validating an MLIP for reactive chemistry.

    \begin{suppinfo}
    First-principles calculations and training data, MLIP architectures and validation, radial distribution functions and dielectric response, collective variables, enhanced sampling and free energy reconstruction, size-dependent ion-count profiles, minimum free energy paths, water wire and hydrogen-bond analyses, and path integral simulation details (PDF)
    \end{suppinfo}

    \section{Author Information}

    \textbf{Corresponding Authors}

    \noindent Juan J. de Pablo. Department of Chemical and Biomolecular Engineering, Tandon School of Engineering, New York University, Brooklyn, New York 11201, USA. Department of Computer Science, Courant Institute of Mathematical Sciences and Department of Physics, New York University, New York City, New York 10012, USA. Email: jjd8110@nyu.edu

    \noindent Yinan Xu. Department of Chemical and Biomolecular Engineering, Tandon School of Engineering, New York University, Brooklyn, New York 11201, USA. Email: yx3878@nyu.edu

    \textbf{Author Contributions}

    \noindent S.V. contributed to design, ideation, code, simulation, analysis, and writing. Y.X. contributed to design, ideation, code, and writing. Y.J. contributed to code, analysis, and writing. G.R.P.-L. contributed to code, analysis, and writing. P.F.Z.R. contributed to code and analysis. Z.Y. contributed to writing. J.M.M., P.F.N., S.B.D., Z.-G.W., and J.J.d.P. contributed to project design, ideation, and editing. All authors have given approval to the final version of the manuscript.

    \textbf{Notes}

    \noindent The authors declare no competing financial interest.

    \section{Data Availability}
    The trained machine learning interatomic potentials (DPSR, deep Wannier, and DPLR), the revPBE-D3 reference training and validation datasets, the simulation input files for DeePMD-kit, LAMMPS, ASE, PySAGES, and i-PI, and the converged free energy profiles and surfaces underlying every figure in this work will be made available upon publication.

    \begin{acknowledgement}
    This research is primarily supported as part of the Advanced Materials for Energy-Water Systems (AMEWS) Energy Frontier Research Center, funded by the U.S. Department of Energy (DOE), Office of Science, Basic Energy Sciences (BES). The development of the advanced sampling methods and PySAGES code employed here is supported by the Midwest Integrated Center for Computational Materials (MICCoM), a DOE Office of Science, Basic Energy Sciences research center. S.V. was supported by the U.S. Department of Energy, Office of Science, Office of Advanced Scientific Computing Research, Department of Energy Computational Science Graduate Fellowship under Award Number DE-SC0022158 during a portion of this work. Y.J. is grateful for a training award from the National Science Foundation under NRT Grant No. 2022023. The authors also acknowledge the computational resources available at Argonne National Laboratory, and at the Research Computing Center at the University of Chicago. Claude-code and OpenAI-Codex agentic AI models were used to aid in coding for both simulation and analysis, as well as to aid in spelling and grammar checks during writing and editing.
    \end{acknowledgement}

    \newpage

\clearpage
\setcounter{figure}{0}
\setcounter{table}{0}
\setcounter{equation}{0}
\setcounter{section}{0}
\renewcommand{\thefigure}{S\arabic{figure}}
\renewcommand{\thetable}{S\arabic{table}}
\renewcommand{\theequation}{S\arabic{equation}}

\begin{center}
{\large\bfseries Supporting Information}\\[4pt]
{\large Fortuitous Cancellation of Competing Effects in the Free Energy of
Water Autoionization: Long-Range Electrostatics, System Size, and Nuclear
Quantum Effects}
\end{center}
\vspace{1em}
\section{First-principles reference calculations}
    
All first-principles calculations were performed using the Quickstep module of CP2K\cite{kuhne_cp2k_2020,lehtola_recent_2018,marek_elpa_2014,marques_libxc_2012} with the revised Perdew-Burke-Ernzerhof generalized gradient approximation functional\cite{perdew_generalized_1996,zhang_comment_1998} combined with Grimme's D3 dispersion correction with zero damping (revPBE-D3(0)).\cite{grimme_consistent_2010,grimme_effect_2011} This functional has been shown to give accurate structural properties for liquid water, with radial distribution functions in good agreement with experiment.\cite{ohto_structure_2019,dodia_structure_2019,ruiz_pestana_ab_2017} Kohn-Sham orbitals were expanded in a triple-$\zeta$ valence basis set with two sets of polarization functions (TZV2P) optimized for condensed-phase calculations,\cite{nagata_surface_2016,li_optimized_2021} and a plane-wave cutoff of 1200 Ry was used for the auxiliary density. Core electrons were treated using Goedecker-Teter-Hutter (GTH) pseudopotentials,\cite{goedecker_separable_1996,krack_pseudopotentials_2005} leaving six valence electrons on each oxygen and one on each hydrogen.

Initial \textit{ab initio} molecular dynamics (AIMD) trajectories were generated using the CP2K calculator within the Atomic Simulation Environment.\cite{hjorth_larsen_atomic_2017} Simulations were performed in the NVT ensemble at 300~K using a Langevin thermostat with a friction coefficient of 0.01~fs$^{-1}$ and a 0.5~fs timestep, on periodic cubic cells containing 128 water molecules with a side length of 15.646~\AA, corresponding to a density of 1.0~g~cm$^{-3}$. Following standard practice, the hydrogen mass was set to 2.014~g~mol$^{-1}$ (heavy water) to permit the larger timestep.

\section{Training data generation}

The MLIP training data were assembled using an active-learning protocol designed to ensure adequate coverage of the autoionization reaction coordinate. Because unbiased AIMD does not sample proton transfer events with sufficient frequency, we used the collective variables defined in the section ``\nameref{si:cvs}'' to bias the system across the autoionization pathway, accumulating configurations via spectral-ABF enhanced sampling as implemented in PySAGES.\cite{zubieta_rico_pysages_2024,zubieta_rico_efficient_2025} To reduce the cost of AIMD-driven exploration, a preliminary MLIP was trained on a small initial dataset and used to rapidly traverse the CV space at MLIP cost. CP2K reference energies and forces were then evaluated on uncorrelated configurations sampled uniformly across the explored regions of phase space.

To improve the generalization of the final MLIPs, the training set was augmented with configurations collected at 450~K and 633~K, and at two system sizes (128 and 256 water molecules in cubic cells of 15.646~\AA{} and 19.713~\AA, respectively). We additionally incorporated a subset of the publicly released configurations of Dasgupta \textit{et al.} \cite{dasgupta_nuclear_2025} that span temperatures from 198~K to 368~K with 256 water molecules, after recomputing energy and forces at the revPBE-D3 level of theory. The final dataset used to train the DPSR and DPLR models contained 17{,}905 training configurations and 4{,}746 validation configurations. The DW model was trained on the smaller subset for which Wannier centers had been computed, comprising 4{,}140 training and 1{,}092 validation configurations.

\begin{figure}[hbt!]
    \centering
    \includegraphics[width=1\linewidth]{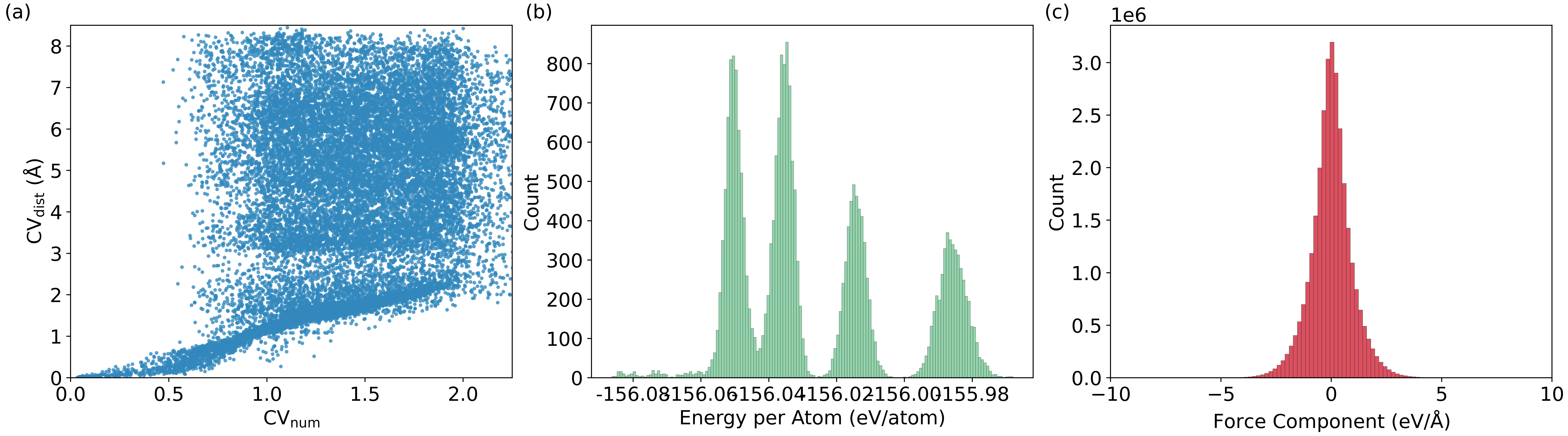}
    \caption{(a) Distribution of training data in CV-space. (b) Distribution of per-atom energies. (c) Distribution of atomic forces (x,y,z included) in training data.}
    \label{fig:training-data}
\end{figure}

The distribution of the training set is illustrated in Figure~\ref{fig:training-data}. The data cover the CV space of interest at several temperatures.

\section{Machine learning interatomic potentials}
\label{si:mlips}

We trained two MLIPs that share the same underlying DFT reference (revPBE-D3) and the same neural-network framework, but differ in their treatment of long-range interactions. Both were trained within the DeePMD-kit framework\cite{wang_deepmd-kit_2018,zeng_deepmd-kit_2023} using the ``se\_e2\_a'' two-body embedding descriptor.\cite{zhang_end--end_2018} The first is a standard short-range model with a finite cutoff radius, which we call the Deep Potential Short Range (DPSR) model. The second, which we call the Deep Potential Long Range (DPLR) model, additionally incorporates explicit long-range electrostatic interactions.\cite{zhang_deep_2022} Architectural details for each network are summarized below, and a comparison of force and energy accuracy across the validation set is provided in the section ``\nameref{app:validation}''. Both models use the same reference data and have similar force errors (Table~\ref{tab:model_errors}), allowing us to compare their treatment of long-range interactions. Their energy errors differ, so this comparison does not isolate the explicit electrostatic term from all other model errors.

\subsection{Deep Potential Short Range (DPSR)}

DPSR follows the Deep Potential Smooth Edition formulation,\cite{zhang_end--end_2018} in which the total energy is written as a sum of atomic contributions,
\begin{equation}\label{eq:edpsr}
    E^{\text{DPSR}} = \sum_i E_i\!\left(\{\mathbf{r}_{ij}\}_{j \in \mathcal{N}_i^{r_c}}\right),
\end{equation}
where each atomic energy $E_i$ depends only on neighbors within a finite cutoff radius $r_c$. The local environment is encoded by a set of symmetry-invariant descriptors and fed into a feedforward network. The model has no explicit Coulomb or dispersion tail beyond the local environment, although the fitted atomic energies can absorb contributions from the reference data. This is the standard form used in most MLIP-based studies of liquid water to date.

The DPSR descriptor network had topology $[32, 64, 128]$ with 16 axis neurons, a neighbor cutoff of 6.0~\AA, and a smoothing onset of 0.5~\AA. These choices are consistent with previous studies of water autoionization.\cite{calegari_andrade_probing_2023,dasgupta_nuclear_2025} The fitting network had three hidden layers of 240 neurons each. The model was trained jointly on energies and atomic forces.

\subsection{Deep Wannier (DW)}

The DW network provides the local electronic polarization needed by DPLR. The DW model predicts the displacement of the Wannier centroid of each oxygen relative to the oxygen nucleus, expressed as a vector-valued (``dipole'') output. Its descriptor network had topology $[16, 32, 64]$ with 16 axis neurons, a cutoff of 6.0~\AA, and a smoothing onset of 0.5~\AA. The fitting network had three hidden layers of 240 neurons each. Each oxygen carries four maximally localized Wannier centers corresponding to two bonding pairs and two lone pairs in neutral water, each center representing two electrons and therefore carrying a charge of $-2e$. The Wannier centroid for a given oxygen is defined as the centroid of its four associated centers, identified as the four Wannier centers closest to the oxygen nucleus. Maximally localized Wannier functions were calculated within CP2K using the Jacobi method with a convergence tolerance of 10$^{-6}$.

\subsection{Deep Potential Long Range (DPLR)}

DPLR augments DPSR with an explicit long-range electrostatic term, following the Deep Potential Long Range framework of Zhang \textit{et al.}\cite{zhang_deep_2022} The total energy takes the form
\begin{equation}\label{eq:edplr}
    E^{\text{DPLR}} = E^{\text{SR}} + E^{\text{LR}}_{\text{Ewald}},
\end{equation}
where the short-range component $E^{\text{SR}}$ is itself a Deep Potential network trained to absorb the difference between the DFT reference and the explicit long-range Ewald term, which is evaluated under periodic boundary conditions. The network fits the residual energy, while the explicit term describes the long-range electrostatics of the assigned charge distribution.

The DPLR model is built on top of the trained DW network. Its short-range descriptor and fitting networks are architecturally identical to those of DPSR, with topology $[32, 64, 128]$, 16 axis neurons, a cutoff of 6.0~\AA, and a smoothing onset of 0.5~\AA{} for the descriptor, and three 240-neuron hidden layers for the fitting network. The explicit long-range electrostatic contribution is constructed by placing Gaussian-smeared point charges of $+6e$ on each oxygen, $+1e$ on each hydrogen, and $-8e$ at the corresponding Wannier centroid predicted by the DW model. These charge assignments are consistent with the valence-electron count exposed by the GTH pseudopotentials used in the underlying DFT reference, giving a net charge of zero for H$_2$O, $+e$ for H$_3$O$^+$, and $-e$ for OH$^-$. The Gaussian smearing parameter was set to $\beta = 0.40$, and the Ewald reciprocal-space grid resolution was set to 1. During inference, the number of gridpoints in K-space was set to give a resolution of 0.5 \AA, as recommended by Zhang et al.\cite{zhang_deep_2022}

\subsection{Molecular dynamics}

Production molecular dynamics was performed with ASE\cite{hjorth_larsen_atomic_2017} for DPSR and with LAMMPS\cite{thompsonLAMMPSFlexibleSimulation2022} for DPLR, in both cases in the NVT ensemble using a Langevin thermostat and a 0.5~fs timestep.

\section{MLIP validation}
\label{app:validation}

We evaluate the MLIPs through several standard benchmarks to ensure that they faithfully reproduce the energies and forces of the reference revPBE-D3 data. Here, we report the energy and force errors of the DPSR and DPLR models (Table~\ref{tab:model_errors}), as well as the error of the DW model in predicting the Wannier centroid positions (Table~\ref{tab:dw_errors}). We also report the errors in specific regions of the two-dimensional CV space (Table~\ref{tab:pathway_model_errors}) to show that the model accuracy remains relatively uniform across the sampled region.

\begin{table}[htbp]
\caption{Energy and force errors for the revPBE-D3 machine-learned models. Energy errors are reported as mean absolute errors (MAE) in meV/atom, and force errors are reported as component-wise root-mean-square errors (RMSE) in meV/\AA.}
\label{tab:model_errors}
\centering
\begin{tabular}{llcc}
\hline
Model & Split & Energy MAE (meV/atom) & Force RMSE (meV/\AA) \\
\hline
DPSR & Training & 0.48 & 38.45 \\
DPSR & Validation & 0.48 & 39.03 \\
DPSR & Testing & 0.49 & 38.91 \\
\hline
DPLR & Training & 0.30 & 36.53 \\
DPLR & Validation & 0.30 & 36.90 \\
DPLR & Testing & 0.30 & 36.70 \\
\hline
\end{tabular}
\end{table}

\begin{table}[htbp]
\caption{Wannier-centroid displacement errors for the revPBE-D3 DW model. Errors are computed over the Cartesian components of the predicted Wannier-centroid displacements and are reported in m\AA.}
\label{tab:dw_errors}
\centering
\begin{tabular}{lcc}
\hline
Split & Displacement MAE (m\AA) & Displacement RMSE (m\AA) \\
\hline
Training & 1.1 & 1.5 \\
Validation & 1.1 & 1.5 \\
\hline
\end{tabular}
\end{table}

\begin{table}[htbp]
\caption{Energy and force errors evaluated at representative points along the revPBE-D3 minimum free energy pathway. The pathway regions are the same as those used in Figure 4 in the main text, with the addition of a neutral state for regular bulk water. Energy errors are MAE in meV/atom, and force errors are component-wise RMSE in meV/\AA.}
\label{tab:pathway_model_errors}
\centering
\begin{tabular}{llrcc}
\hline
Model & Region & Frames & Energy MAE (meV/atom) & Force RMSE (meV/\AA) \\
\hline
DPSR & Neutral & 1524 & 0.44 & 33.49 \\
DPSR & WD & 337 & 0.37 & 36.20 \\
DPSR & CIP & 89 & 0.38 & 41.55 \\
DPSR & TS & 58 & 0.42 & 41.84 \\
DPSR & SSIP & 124 & 0.51 & 40.60 \\
DPSR & TS2 & 119 & 0.64 & 41.02 \\
DPSR & WS & 125 & 0.69 & 41.39 \\
\hline
DPLR & Neutral & 1524 & 0.23 & 34.69 \\
DPLR & WD & 337 & 0.30 & 36.58 \\
DPLR & CIP & 89 & 0.31 & 41.24 \\
DPLR & TS & 58 & 0.32 & 41.64 \\
DPLR & SSIP & 124 & 0.29 & 40.45 \\
DPLR & TS2 & 119 & 0.34 & 41.16 \\
DPLR & WS & 125 & 0.27 & 41.83 \\
\hline
\end{tabular}
\end{table}

\clearpage
\subsection{Deep Potential Short Range (DPSR)}

\begin{figure}[hb!]
    \centering
    \includegraphics[width=0.9\linewidth]{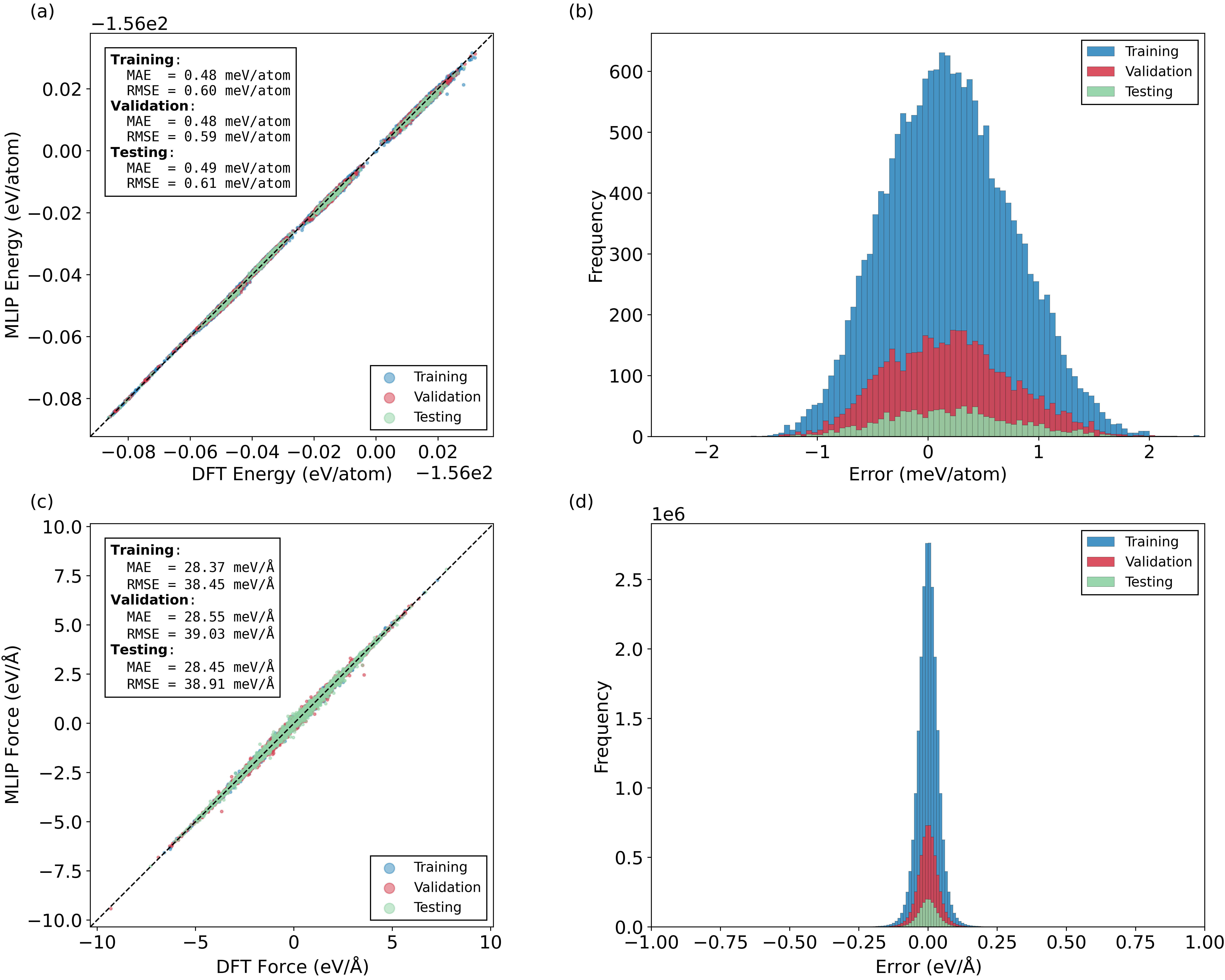}
    \caption{Performance metrics of DPSR model trained on revPBE-D3 reference data. (a) Parity plot of potential energy per atom, showing an MAE of 0.49 meV/atom for the testing set. (b) The distribution of errors in per-atom energies across datasets. (c) Parity plot of force, showing an RMSE of 38.91 meV/{\AA} for the testing set. (d) The distribution of errors in the force components across datasets.}
    \label{fig:dpsr-error}
\end{figure}

The performance of the optimized DPSR MLIP is presented in Figure \ref{fig:dpsr-error}, which displays the error in per-atom potential energies and forces. Our DPSR model achieves an MAE in energy of 0.49 meV/atom on the testing set. Furthermore, the RMSE in the force components on the testing set is 38.91 meV/{\AA}.

\begin{figure}[hbt!]
    \centering
    \includegraphics[width=1\linewidth]{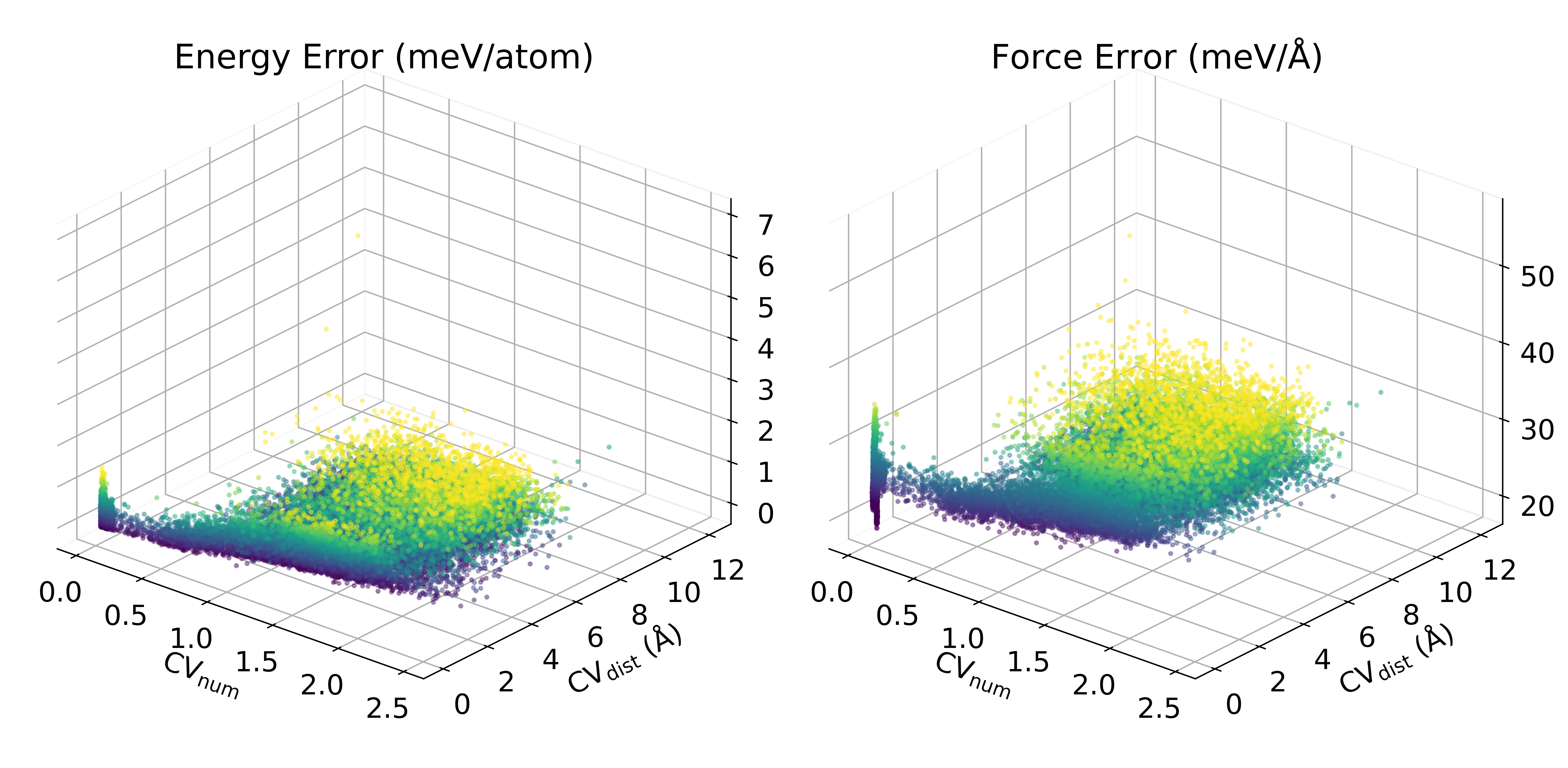}
    \caption{Distribution of errors in per-atom energies (left) and forces (right) across the relevant CV space for the trained DPSR model.}
    \label{fig:dpsr-cv-error}
\end{figure}

Since we aim to conduct enhanced sampling simulations in order to probe the reaction free energy, it is important that the MLIP is accurate across the whole range of the collective variable space that we intend to sample. The error distributions in Figure \ref{fig:dpsr-cv-error} demonstrate that the MLIP maintains uniform accuracy throughout the relevant CV space in terms of both energies and forces.

\clearpage
\subsection{Deep Potential Long Range (DPLR)}

\begin{figure}[hbt!]
    \centering
    \includegraphics[width=0.9\linewidth]{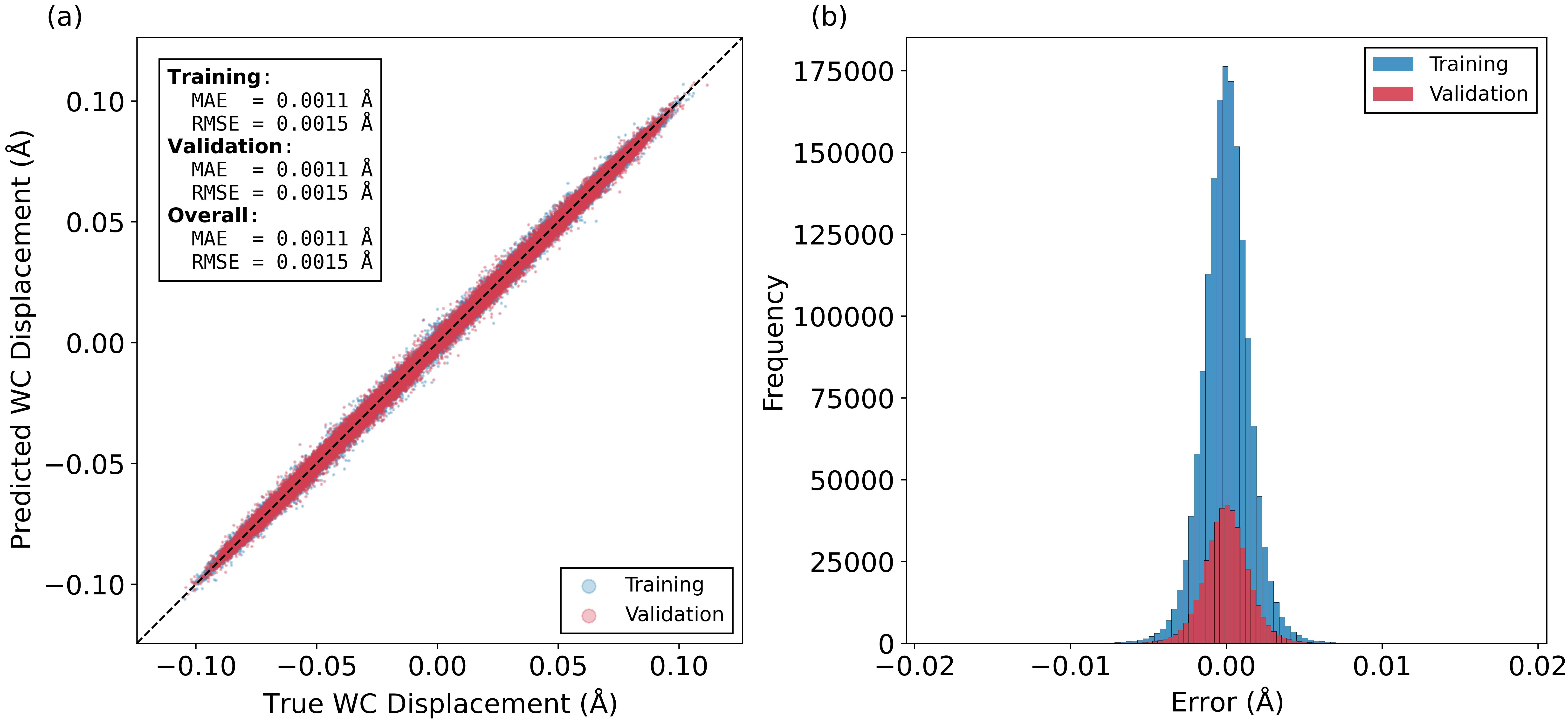}
    \caption{Performance metrics of DW model trained on revPBE-D3 reference data. (a) Parity plot of Wannier Centroid (WC) displacement between the trained model and the reference revPBE-D3 theory, showing a validation RMSE of 1.5~m\AA. (b) The distribution of errors.}
    \label{fig:wannier-parity}
\end{figure}

The DPLR model requires first training a lightweight model to predict the displacement of the Wannier centroid from each oxygen atom, sometimes denoted a deep Wannier (DW) model.\cite{zhang_deep_2022} In Figure \ref{fig:wannier-parity}, we provide the error in the predicted displacement of the Wannier centroid by our trained DW model.

\begin{figure}[hbt!]
    \centering
    \includegraphics[width=1\linewidth]{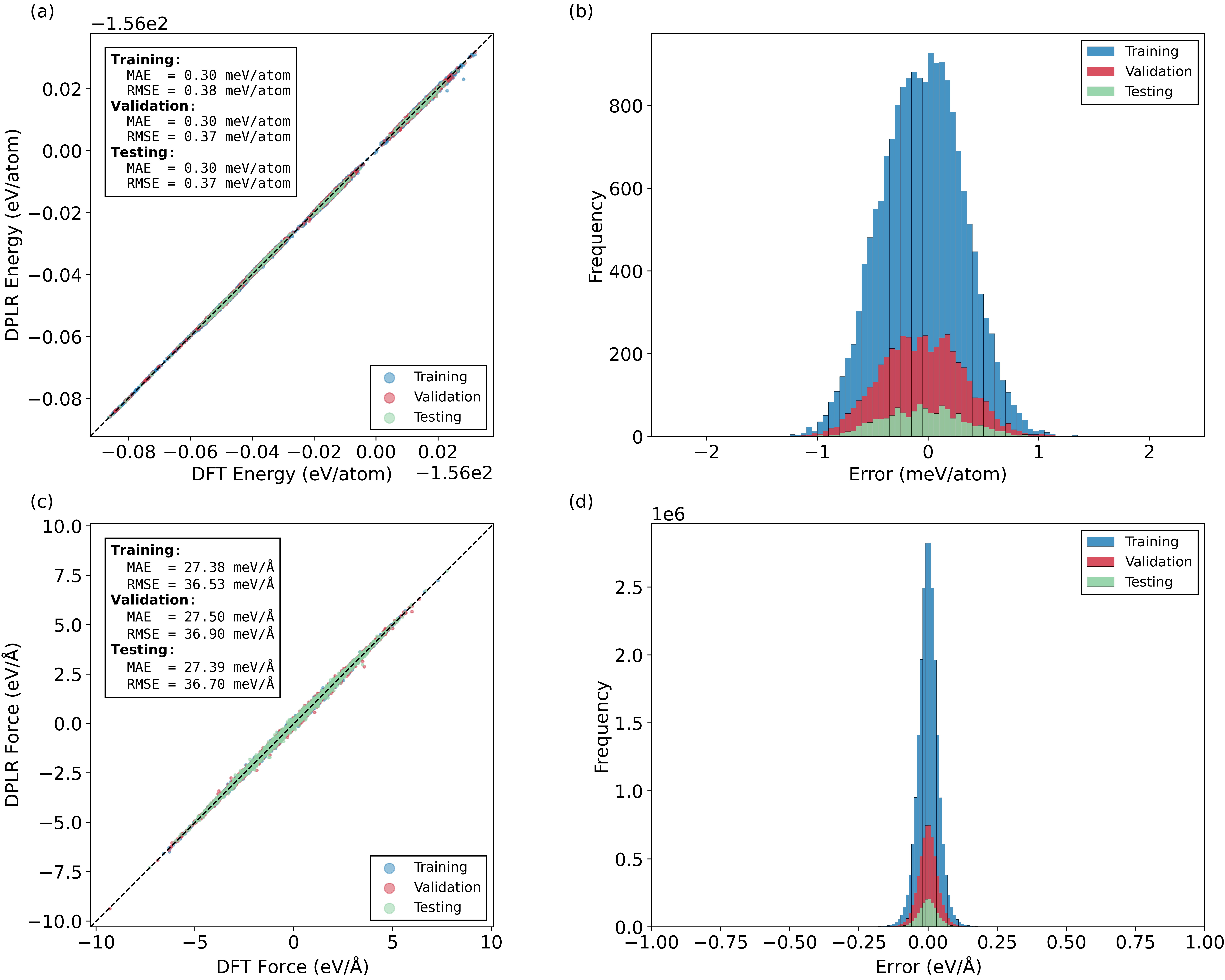}
    \caption{Performance metrics of DPLR model trained on revPBE-D3 reference data. (a) Parity plot of potential energy per atom, showing an MAE of 0.30 meV/atom for the testing set. (b) The distribution of errors in per-atom energies across datasets. (c) Parity plot of force, showing an RMSE of 36.70 meV/{\AA} for the testing set. (d) The distribution of errors in the force components across datasets.}
    \label{fig:dplr-error}
\end{figure}

Since the DPLR model trained on top of the DW model incorporates long-ranged interactions beyond the cutoff of the descriptor network, it is expected that this model will be more accurate at predicting energy and forces, particularly when dipoles and charges are present in the system. We indeed observe a measurable increase in energy accuracy, as shown in Figure \ref{fig:dplr-error}. On the testing set, the MAE in per-atom potential energies is 0.30 meV/atom, compared to 0.49 meV/atom for the DPSR model. The force RMSEs are 36.70 and 38.91 meV/{\AA} for DPLR and DPSR, respectively. Despite their similar force errors, the models give substantially different autoionization free energies, as shown in the main text. Force and energy validation alone therefore does not establish the accuracy of the reaction free energy.

\begin{figure}[hbt!]
    \centering
    \includegraphics[width=1\linewidth]{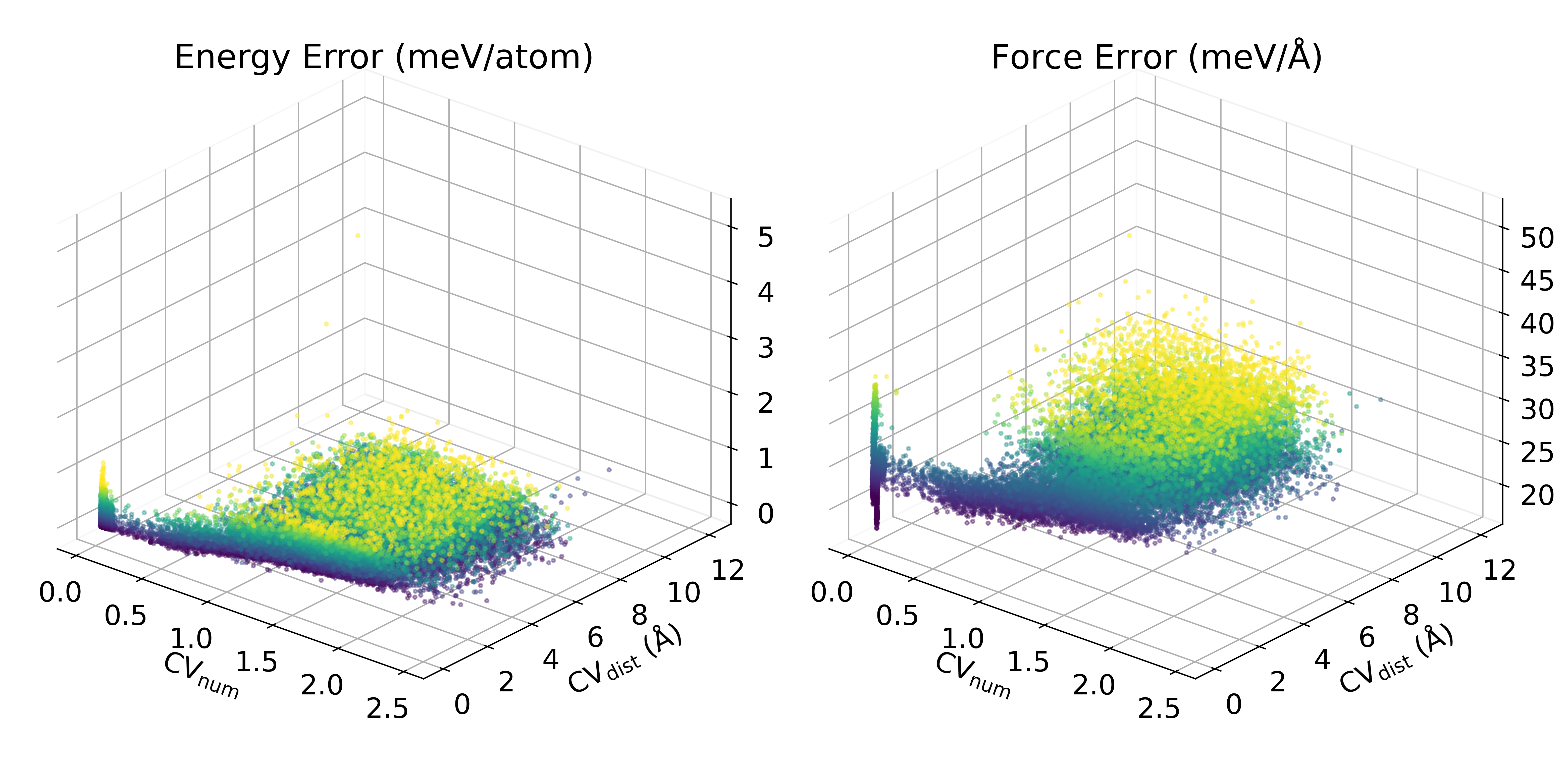}
    \caption{Distribution of errors in per-atom energies (left) and forces (right) across the relevant CV space for the trained DPLR model.}
    \label{fig:dplr-cv-error}
\end{figure}

Again, it is important that the potential is not only accurate on average, but also that it is uniformly accurate across the collective variable space of interest. To this end, we provide the errors as a function of collective variable in Figure \ref{fig:dplr-cv-error}.

\clearpage
\subsection{Model performance on collective properties}

\begin{figure}[hbt!]
    \centering
    \includegraphics[width=1\linewidth]{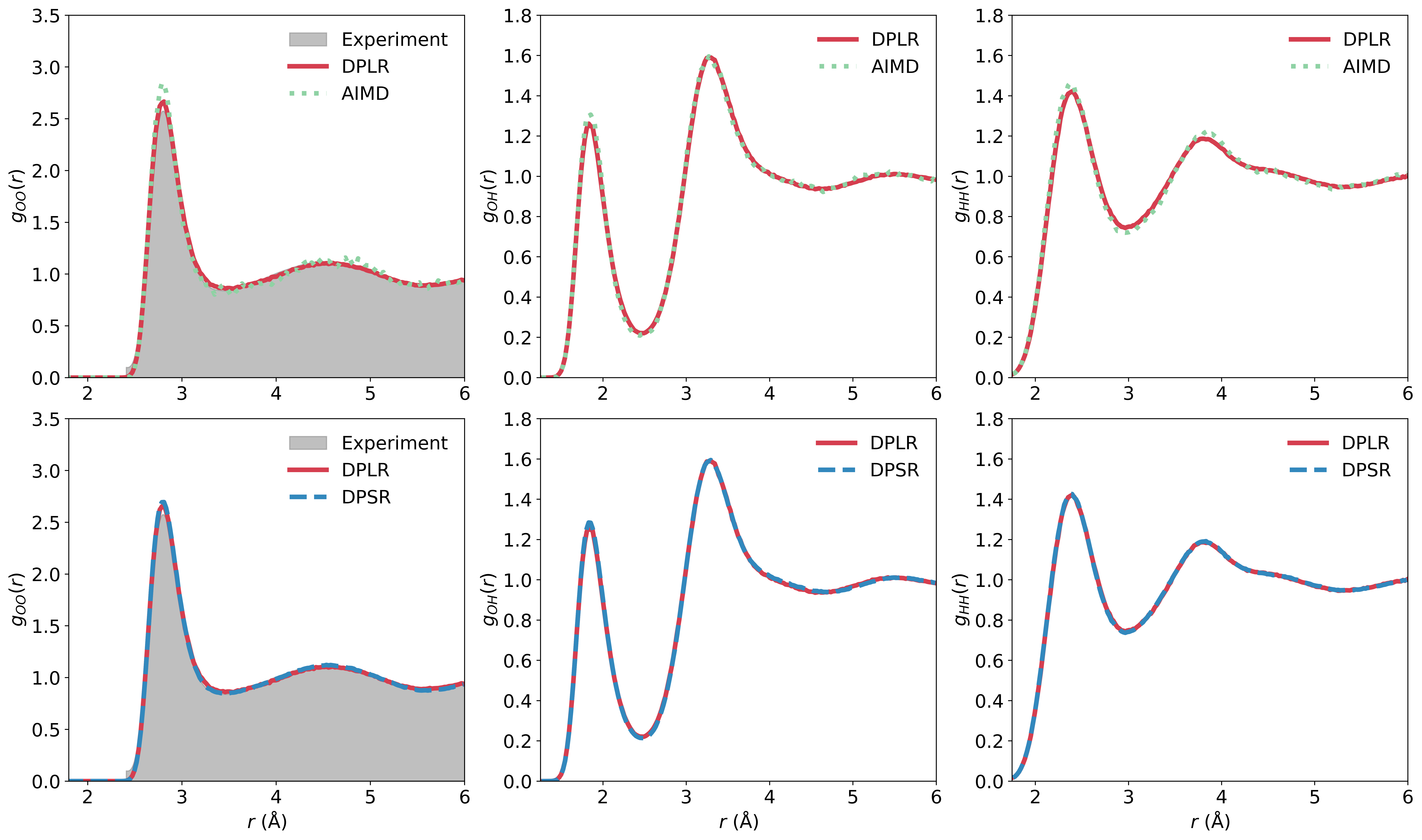}
    \caption{Radial (pair) distribution functions for OO, OH, and HH computed from MD simulations using the trained DPLR model, DPSR model, and the reference DFT (revPBE-D3). The experimental data were reproduced from Skinner et al.\cite{skinner_benchmark_2013}}
    \label{fig:rdfs}
\end{figure}

We also compare the models through structural and dielectric properties. First, in Figure \ref{fig:rdfs} we present the radial (pair) distribution functions from MD simulations using DPLR, DPSR, and the revPBE-D3 reference. These results reaffirm that the revPBE-D3 level of theory can accurately reproduce the structure of real water, and also that both the DPSR and DPLR models can faithfully reproduce the structure obtained from DFT.

\begin{figure}[hbt!]
    \centering
    \includegraphics[width=1.0\linewidth]{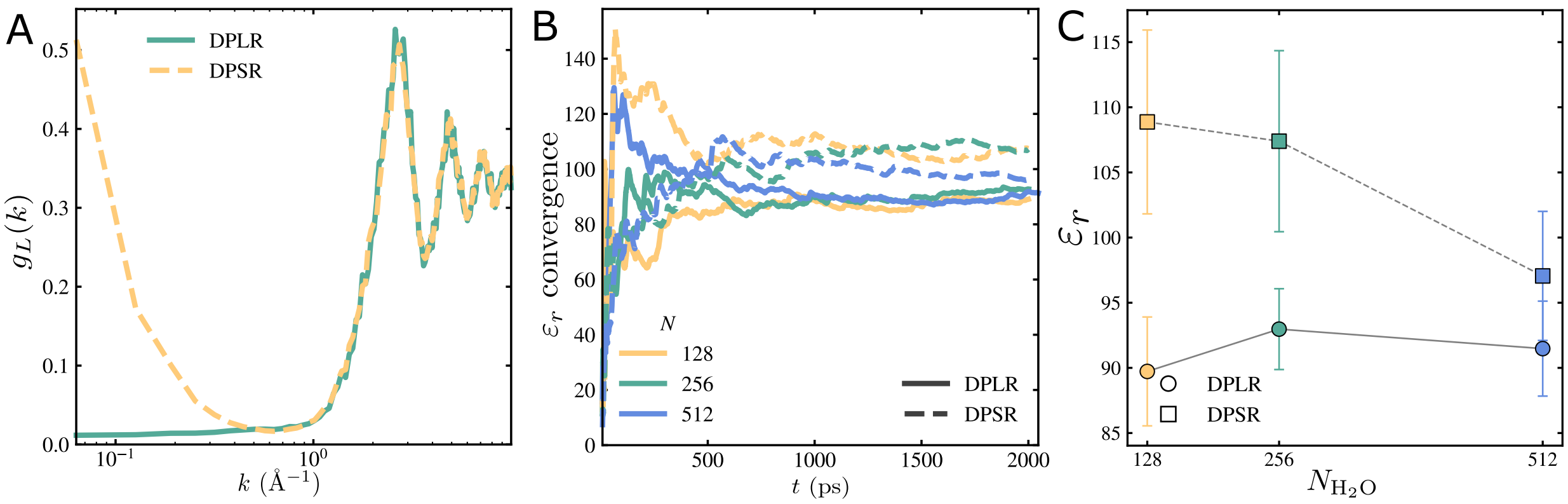}
    \caption{Dielectric response of the short-range (DPSR) and long-range (DPLR) MLIPs. (A) Longitudinal dipole-dipole correlation function $g_L(k)$ computed in an elongated cell (2048 water molecules, $L_z = 99.3$~\AA) with $\mathbf{k}$ along the long axis (DPLR solid, DPSR dashed). (B) Running estimate of the dielectric constant $\varepsilon_r$ from the Neumann fluctuation formula for cubic cells of 128, 256, and 512 water molecules (colors indicate system size, DPLR solid, DPSR dashed). (C) Converged dielectric constant as a function of system size (circles: DPLR, squares: DPSR). Error bars denote the standard error of the mean over 10 trajectory blocks.}
    \label{fig:dielectric}
\end{figure}

We also compute the dielectric constant using both the DPSR and DPLR models across systems containing 128, 256, and 512 waters. The simulations are conducted in the NVT ensemble with a Langevin integrator, $T=300$\,K, deuterated water, $dt=0.5$\,fs, and a total time of 2\,ns. For both the DPSR and DPLR simulations, the simulation frames are post-processed using the Deep Wannier model to obtain the Wannier centroids. The dipole of each water molecule is then determined by assigning the Wannier centroid a charge of -8e, and the O and H atoms charges of +6e and +1e, respectively. The dipoles are then used to compute the system dipole as a function of time, and the dielectric constant is calculated using the Neumann fluctuation formula under conducting (tin-foil) boundary conditions,
$$\varepsilon_r = 1 + \frac{\langle \mathbf{M}^2\rangle - \langle \mathbf{M}\rangle^2}{3\varepsilon_0 V k_B T}, \qquad \mathbf{M}(t) = \sum_{i=1}^{N} \boldsymbol{\mu}_i(t)$$
where $\varepsilon_r$ is the relative dielectric constant, $\mathbf{M}$ is the system dipole, $\varepsilon_0$ is the vacuum permittivity, $V$ is the system volume, $k_B$ is Boltzmann's constant, $T$ is the temperature, and $\boldsymbol{\mu}_i$ is the molecular dipole of molecule $i$. The running estimate of $\varepsilon_r$ converges within the 2\,ns trajectories for all systems (Figure~\ref{fig:dielectric}B). The converged values are shown as a function of system size in Figure~\ref{fig:dielectric}C: the DPLR model yields $\varepsilon_r \approx 90$ with little apparent size dependence, whereas the DPSR model systematically overestimates the dielectric constant relative to DPLR and appears to decrease with system size.

While the dielectric constant characterizes the response at zero wavevector, the two models are best distinguished by resolving the wavevector dependence of the polarization fluctuations. To this end, we compute the longitudinal dipole-dipole correlation function (Figure~\ref{fig:dielectric}A),
$$g_L(k) = \frac{\left\langle \left|\hat{\mathbf{k}}\cdot\mathbf{m}(\mathbf{k})\right|^2 \right\rangle}{N\langle\mu^2\rangle}, \qquad \mathbf{m}(\mathbf{k}) = \sum_{i=1}^{N}\boldsymbol{\mu}_i\, e^{i\mathbf{k}\cdot\mathbf{r}_i}$$
where $\hat{\mathbf{k}}$ is the unit vector along the wavevector $\mathbf{k}$, $\mathbf{r}_i$ is the position of molecule $i$, and $\langle\mu^2\rangle$ is the mean-square molecular dipole. That is, $g_L(k)$ measures the fluctuations of the component of the dipole density parallel to $\mathbf{k}$, and is related to the longitudinal $k$-dependent dielectric function through
$$1 - \frac{1}{\varepsilon_L(k)} = \frac{N\langle\mu^2\rangle}{\varepsilon_0 V k_B T}\, g_L(k).$$
Because $\varepsilon_L(k)$ is finite, continuum electrostatics requires $g_L(k)$ to remain bounded as $k \to 0$, with its limit fixed by $1-1/\varepsilon_L$. To access small wavevectors, these simulations use an elongated cell (2048 H$_2$O, $L_z = 99.3$\,\AA) with $\mathbf{k}$ taken along the long axis, reaching $k_\mathrm{min} = 2\pi/L_z = 0.063$\,\AA$^{-1}$. All other simulation and post-processing details are identical to those described above. As shown in Figure~\ref{fig:dielectric}A, the two models are indistinguishable for $k \gtrsim 0.5$\,\AA$^{-1}$, reflecting their common short-range structure. As $k \to 0$, however, DPLR plateaus at $g_L \approx 0.012$, in quantitative agreement with the value $\left(1-1/\varepsilon_r\right)\varepsilon_0 V k_B T/\left(N\langle\mu^2\rangle\right)$ implied by its dielectric constant, whereas DPSR exhibits a spurious low-$k$ upturn. Lacking the explicit $1/r$ Coulomb interaction, the short-range model fails to suppress long-wavelength longitudinal polarization fluctuations, providing a direct, structural signature of the missing long-range electrostatics. The larger dielectric constant of DPSR suggests stronger screening of the separating ions. This may contribute to its lower separation free energy, although the present comparison does not quantify the contribution.

\clearpage
\section{Collective variables}
\label{si:cvs}

The autoionization free energy is computed using enhanced sampling along two collective variables (CVs) that resolve, respectively, the ionization state of the system and the spatial separation between the resulting ions.\cite{grifoni_microscopic_2019,liu_mechanistic_2023}

The first CV, $\text{CV}_\text{num}$, measures the effective number of autoions present. For each oxygen $i$ we evaluate a continuous coordination number $n_i(\mathbf{R})$ counting nearby hydrogens, and sum the squared deviation from the neutral value of two:
\begin{equation}\label{eq:cvnum}
    \text{CV}_{\text{num}} = \sum_{i=1}^{N_O} \left[ n_i(\mathbf{R}) - 2 \right]^2, \qquad n_i(\mathbf{R}) = \sum_{j=1}^{N_H} \frac{1 - (R_{ij}/R_0)^{16}}{1 - (R_{ij}/R_0)^{56}},
\end{equation}
where $\mathbf{R}$ denotes the complete atomic configuration, $R_{ij}$ is the distance between oxygen $i$ and hydrogen $j$, and $R_0 = 1.32$~\AA{} is a reference distance for the switching function. The 16/56 exponents produce a sharp but smooth switch between bonded ($n_{ij} \to 1$) and nonbonded ($n_{ij} \to 0$) limits, consistent with previous applications of this CV.\cite{liu_mechanistic_2023} In the integer-coordination limit, an intact H$_2$O contributes zero and an OH$^-$ or H$_3$O$^+$ contributes one unit. The smooth switching functions give a nonzero baseline in neutral water, which is subtracted in plots of $\Delta\text{CV}_\text{num}$.

The second CV, $\text{CV}_\text{dist}$, measures the effective separation between the autoions:
\begin{equation}\label{eq:cvdist}
    \text{CV}_{\text{dist}} = -\sum_{i<k}^{N_O} q_i q_k R_{ik}, \qquad q_i = \sum_{j=1}^{N_H} \frac{e^{-\lambda R_{ij}}}{\sum_{i'=1}^{N_O} e^{-\lambda R_{i'j}}} - 2,
\end{equation}
where $q_i$ is a continuous effective charge for oxygen $i$. The soft-min assignment with $\lambda = 8$~\AA$^{-1}$ associates each hydrogen with its nearest oxygen in a smooth and differentiable manner, so that $q_i$ approaches $-1$ for OH$^-$, $0$ for H$_2$O, and $+1$ for H$_3$O$^+$. The subtraction of two places a normally coordinated water at $q_i \approx 0$, so that positive $q_i$ reflects hydronium character and negative $q_i$ hydroxide character. For a single pair with charges $+1$ and $-1$ and all other $q_i=0$, $\text{CV}_\text{dist}$ equals the ion separation. It vanishes when all $q_i=0$, but smooth proton assignment gives small fluctuations in neutral water. The same $q_i$ is used throughout this work to identify the reactive oxygens (the section ``\nameref{sec:water-wire}''), where $Q_\text{min}$ and $Q_\text{max}$ denote its minimum and maximum over all oxygens in a given configuration.

\clearpage
\section{Enhanced sampling and free energy surface construction}
\label{si:sampling}

\subsection{Spectral-ABF and the one- and two-dimensional regions}

We sample along these CVs using the spectral adaptive biasing force method (spectral-ABF)\cite{zubieta_rico_efficient_2025} as implemented in the PySAGES package.\cite{zubieta_rico_pysages_2024} Spectral-ABF represents the time-dependent bias as a truncated cosine series, providing a smooth and rapidly converging estimate of the free energy gradient relative to the histogram-based representations used in standard ABF. Because $\text{CV}_\text{num}$ and $\text{CV}_\text{dist}$ are strongly correlated near neutral water but become effectively independent once two ions have formed, we split the free energy calculation into two regions and sample each with the most appropriate dimensionality. In the region $\text{CV}_\text{num} < 2$, the initial proton transfer is dominated by O--H bond stretching, and we apply one-dimensional ABF along $\text{CV}_\text{num}$. In the region $\text{CV}_\text{num} \gtrsim 1.5$, which includes ion formation and subsequent separation, we apply two-dimensional spectral-ABF along $(\text{CV}_\text{num}, \text{CV}_\text{dist})$ to resolve the free energy as a function of both the ionization state and the ion separation. The two regions overlap sufficiently for the corresponding profiles to be combined into a single free energy along $\text{CV}_\text{dist}$ that covers the entire reaction.

\subsection{Window decomposition and surface reconstruction}

For complex systems, even with the help of enhanced sampling, exploring the CV space can be time-consuming and the convergence can be challenging to achieve. To improve the efficiency, we divide the CV space into multiple simulation windows, each exploring a different region of the free energy surface. While this method enables thorough sampling, it can lead to misalignment when reconstructing the entire energy landscape. To resolve this issue, we developed a mechanism to combine the free energy surface estimation through neural network fitting. We adopted the same spectral function family as in spectral-ABF, using a spectral neural network. Our model employs a three-layer neural network, each layer containing 32 neurons with a tanh activation function. We fit the gradient of the neural network with respect to the CV to the estimated mean force. This ensures that, similar to the spectral-ABF model, the output of the neural network represents the free energy associated with the specified CV, and its gradient corresponds to the estimated mean force. The loss function is given by the following:
\[
\min_\theta\mathcal{L} = \langle\|f'_\theta(x) - \bar F(x)\|_2^2\rangle_x
\]
where $\mathcal{L}$ denotes the loss function, $f'_\theta(x)$ represents the gradient of the model output with respect to the collective variables parameterized by $\theta$, $\bar F(x)$ is the mean force.

\subsection{Reweighting and joining the two regions}

\begin{figure}[hbt!]
    \centering
    \includegraphics[width=1\linewidth]{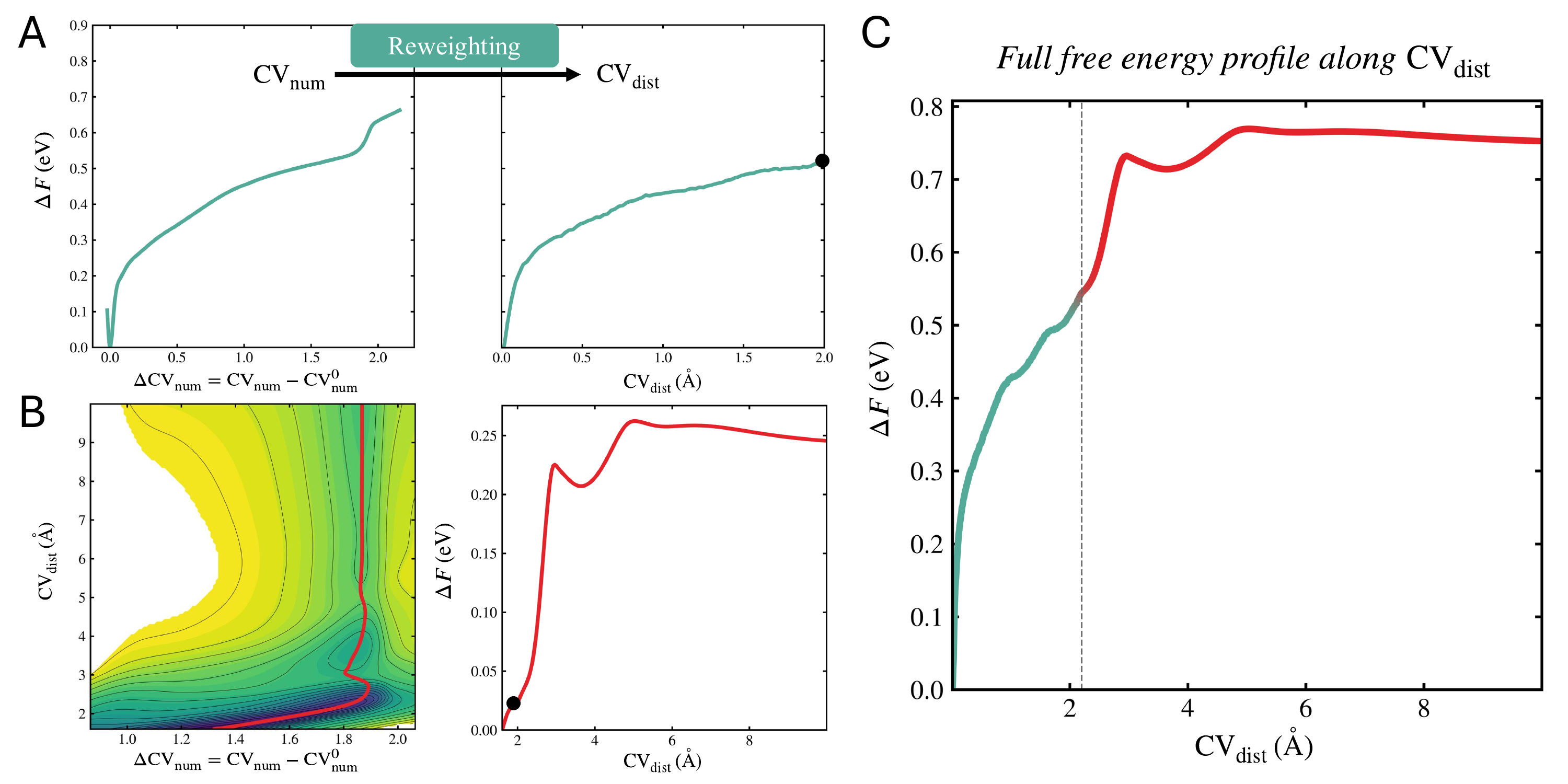}
    \caption{Construction of the full free energy profile along $\text{CV}_\text{dist}$ by combining the one- and two-dimensional sampling regions, computed with the DPLR model and 512 water molecules. (A) The free energy profile from one-dimensional sampling in $\text{CV}_\text{num}$ is reweighted onto $\text{CV}_\text{dist}$ over the region $0 < \text{CV}_\text{dist} < 2$~\AA. (B) The minimum free energy path from two-dimensional sampling in the ionized region is projected along $\text{CV}_\text{dist}$ for $\text{CV}_\text{dist} > 2$~\AA. (C) The two profiles are joined at $\text{CV}_\text{dist} \approx 2$~\AA.}
    \label{fig:reweighting}
\end{figure}

To merge the one- and two-dimensional regions onto a common axis, we reweight the converged one-dimensional profile from $\text{CV}_\text{num}$ onto $\text{CV}_\text{dist}$ using a two-dimensional histogram accumulated during an extended biased run in which both CVs are tracked. The reweighted profile is then joined to the two-dimensional MFEP projected onto $\text{CV}_\text{dist}$ at the overlap point $\text{CV}_\text{dist} \approx 2$~\AA. The construction is illustrated in Figure~\ref{fig:reweighting}. The resulting full profile is integrated as described in the main text to obtain the autoionization free energy and $\text{p}K_w$.

\subsection{Simulated systems}

Free energy profiles were calculated for cubic boxes of 128, 256, and 512 water molecules, with the cell length adjusted to maintain a density of 1.0~g~cm$^{-3}$. To study the effect of temperature, simulations were carried out with the DPLR model and 512 water molecules at 300~K, 450~K, and 600~K, all at a fixed density of 1.0~g~cm$^{-3}$.

\subsection{One-dimensional profiles along $\text{CV}_\text{num}$}

\begin{figure}[hbt!]
    \centering
    \includegraphics[width=0.55\linewidth]{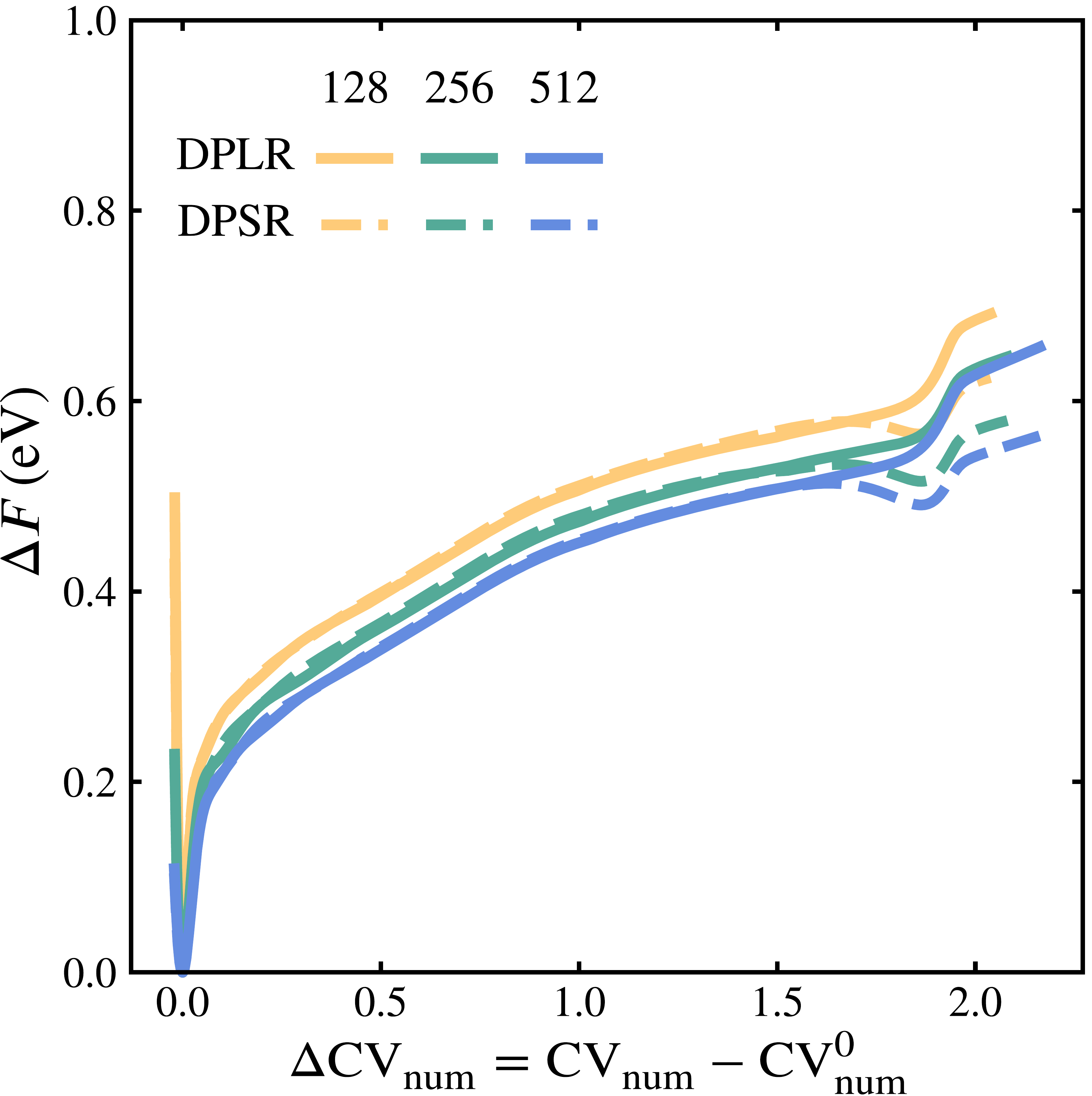}
    \caption{One-dimensional free energy profiles along $\text{CV}_\text{num}$ for 128, 256, and 512 water molecules, computed with the short-range (DPSR, dashed) and long-range (DPLR, solid) models.}
    \label{fig:size-cvnum}
\end{figure}

Figure~\ref{fig:size-cvnum} shows the one-dimensional free energy profiles along $\text{CV}_\text{num}$ for 128, 256, and 512 water molecules with both models. The free energy decreases with system size in a similar manner for DPSR and DPLR, and the two are nearly indistinguishable until $\text{CV}_\text{num}$ approaches the two-ion state, as expected while no charged species are yet present and the participating waters remain close together. The DPSR curves develop a small minimum at the two-ion state that is absent in DPLR. This minimum arises from the decrease in free energy with increasing $\text{CV}_\text{dist}$ that DPSR also experiences but cannot resolve when only $\text{CV}_\text{num}$ is biased.

\section{Minimum free energy path and transition states}

The minimum free energy pathway (MFEP) through the two-dimensional free energy surface is computed using the zero-temperature string method,\cite{e_string_2002} which iteratively relaxes a discrete chain of states onto the steepest-descent path connecting the contact ion pair and the fully separated state. The method was implemented in a Python script with NumPy and PyTorch, using the combined PyTorch free energy surface generator model of the section ``\nameref{si:sampling}'' to evaluate the free energy and its gradient at arbitrary points in the CV space.

The initial and final states are determined by relaxing the endpoints in the CV space, and the string is initialized by linear interpolation between them. At each iteration every image is displaced along the local free energy gradient and the string is then reparameterized to restore uniform spacing along its arc length, which prevents the images from sliding into the endpoint basins and keeps the discretization even. Additional images are inserted dynamically when the free energy difference between neighboring images exceeds 0.05~eV or their separation in the CV space exceeds 0.1. Iteration continues until the string no longer moves appreciably, and the converged MFEP is obtained by interpolating between the final images. The transition states reported in the main text are the free energy maxima along this converged string.

\section{Water wire distribution analysis}
\label{sec:water-wire}
The water wire connecting the incipient hydroxide and hydronium ions was characterized directly from the enhanced sampling trajectories. For every frame logged from the two-dimensional spectral-ABF windows, we evaluated a set of graph-based descriptors that identify the two reactive oxygens and trace the shortest hydrogen-bonded path between them. The descriptors use smooth switching functions and were evaluated under the minimum-image convention. Selection of the reactive oxygens and shortest path can introduce nondifferentiable points when the selected atoms or path change.

\subsection{Identification of the reactive oxygens}
The reactive oxygens are identified using the same effective charge $q_i$ that enters $\text{CV}_\text{dist}$, defined in Equation~\ref{eq:cvdist}, which measures each oxygen's excess or deficit of covalently associated protons. The oxygens with the minimum and maximum $q_i$ define the hydroxide-like and hydronium-like endpoints of the wire. Their effective charges are denoted $Q_\text{min}$ and $Q_\text{max}$.

\subsection{Hydrogen-bond network and shortest path}
We next build a hydrogen-bond adjacency between every pair of oxygens $(i,j)$ as the product of a distance and an angular switching function,
\begin{equation}
    h_{ij} = \sigma\!\left[\beta_d\,(R_c - R_{ij})\right]\,
             \max_{k}\Big\{ \sigma\!\left[\beta_a\,(\theta_{ikj} - \theta_c)\right]\, g_{ikj}\Big\},
    \label{eq:hbond}
\end{equation}
where $\sigma(x) = (1+e^{-x})^{-1}$ is the logistic sigmoid, $R_{ij}$ is the O--O distance, and $\theta_{ikj}$ is the O$_i$--H$_k$--O$_j$ angle subtended at the bridging hydrogen $k$. The maximum over $k$ selects the best mediating hydrogen, and the term $g_{ikj} = \max\{\sigma[\beta_d(R_c - r_{ik})],\, \sigma[\beta_d(R_c - r_{jk})]\}$ requires that hydrogen to lie within $R_c$ of at least one of the two oxygens. We use an O--O cutoff of $R_c = 3.7$~\AA{} and an O--H--O angular cutoff of $\theta_c = 130^{\circ}$, with sharpness parameters $\beta_d = 15$~\AA$^{-1}$ and $\beta_a = 0.5$~deg$^{-1}$.

Each frame is then represented as a graph whose nodes are oxygens and whose edges are the identified hydrogen bonds. Edges are gated with a steep sigmoid $\gamma_{ij} = \sigma[50\,(h_{ij} - \tfrac{1}{2})]$ centered at $h_{ij} = \tfrac{1}{2}$, and each edge is assigned a weight $W_{ij} = R_{ij}\,\gamma_{ij} + \Lambda\,(1 - \gamma_{ij})$, where $\Lambda$ is a large sentinel that effectively removes non-bonded pairs. The water wire is the shortest path from the $Q_\text{min}$ oxygen to the $Q_\text{max}$ oxygen, computed with the Bellman-Ford algorithm in min-plus algebra. The wire length is the sum of O--O distances along this path, and the number of members is the number of oxygens it contains, obtained by repeating the shortest-path calculation with unit edge weights and adding one to count nodes rather than edges. Because the wire follows the hydrogen-bond network rather than a straight line, its length is at least the direct $Q_\text{min}$--$Q_\text{max}$ separation, with equality for two-membered wires and, in principle, for longer collinear paths.

\subsection{Reweighting and state definitions}
The wire descriptors were pooled across all $16$ two-dimensional sampling windows. To recover unbiased statistics, each frame $a$ was assigned a Boltzmann weight
\begin{equation}
    w_a = \frac{e^{-F(\mathbf{s}_a)/k_BT}}{\sum_b e^{-F(\mathbf{s}_b)/k_BT}},
    \label{eq:wire-weight}
\end{equation}
where $\mathbf{s}_a = (\text{CV}_\text{num}, \text{CV}_\text{dist})$ are the biased collective variables of the frame, $F$ is the converged two-dimensional free energy surface interpolated at $\mathbf{s}_a$, $k_B$ is the Boltzmann constant, and $T = 300$~K. The characteristic states along the pathway (WD, CIP, TS, SSIP, TS2, and WS) were defined by elliptical regions in the CV plane,
\begin{equation}
    \frac{(\text{CV}_\text{num} - c_\text{num})^2}{\Delta_\text{num}^2}
    + \frac{(\text{CV}_\text{dist} - c_\text{dist})^2}{\Delta_\text{dist}^2} \leq 1,
    \label{eq:wire-window}
\end{equation}
centered on each state $(c_\text{num}, c_\text{dist})$ with semi-axes $\Delta_\text{num} = 0.06$ and $\Delta_\text{dist} = 0.26$~\AA. For each state we then computed weighted kernel density estimates of the wire length, resolved by the number of members, which are shown as the violin plots in Figure 4A of the main text. The width of each violin reflects the relative weighted frequency of a given wire size at that state, and its vertical extent reflects the range of wire lengths observed.

\newpage
\section{Hydrogen bond coordination number analysis}
We report the hydrogen bond coordination number (HBCN) at the same characteristic states (WD, CIP, TS, SSIP, TS2, and WS) used in the water wire analysis (the section ``\nameref{sec:water-wire}''), using the identical elliptical CV windows and Boltzmann reweighting. Following \citet{liu_mechanistic_2023}, we calculate HBCN for the $i$-th oxygen atom in the system as
\begin{equation}
HBCN_i = \sum_{k=1,\ k\neq i}^{n_O}\left[\frac{1-(R_{ik}/R_1)^{16}}{1-(R_{ik}/R_1)^{56}}\right]\cdot\left[\frac{1-(\theta_{ik}/\theta_1)^{16}}{1-(\theta_{ik}/\theta_1)^{56}}\right]
\end{equation}
where $R_{ik}$ is the distance between oxygens $i$ and $k$, and $\theta_{ik}$ is the supplement of the O-H-O angle at the bridging hydrogen. We use $R_1 = 3.3$~\AA{} and $\theta_1 = 45^{\circ}$.

\newpage
\section{Nuclear Quantum Effects}
We include nuclear quantum effects using path integral molecular dynamics (PIMD). Machine learning interatomic potentials reduce the cost of force evaluation, while colored-noise thermostats accelerate convergence with respect to the number of path integral beads.\cite{ceriotti_accelerating_2011,ceriotti_efficient_2012}

We performed PIMD simulations using the i-PI engine\cite{kapil_i-pi_2019} coupled to LAMMPS\cite{thompsonLAMMPSFlexibleSimulation2022} for force evaluation with the DPLR model. To reduce the computational overhead, we employed the Path-Integral Generalized Langevin Equation (PIGLET) methodology, which utilizes frequency-dependent colored noise to accelerate the convergence of quantum observables.\cite{ceriotti_accelerating_2011,ceriotti_efficient_2012} We used $n_\text{beads}=8$. Simulations were performed on cubic boxes of 512 H$_2$O or D$_2$O molecules at 300~K and a density of 1.0~g~cm$^{-3}$. Timesteps of 0.2~fs and 0.5~fs were used for H$_2$O and D$_2$O, respectively.

To ensure a direct comparison with our classical results, we applied the same spectral-ABF enhanced sampling protocol\cite{zubieta_rico_efficient_2025} within the PySAGES suite\cite{zubieta_rico_pysages_2024} used for the classical simulations to compute the one-dimensional free energy profiles, two-dimensional free energy surfaces (FES), and the resulting $pK_w$ for autoionization.

\newpage

\clearpage
\bibliography{references}

\end{document}